\documentclass[aps,prb,preprint,longbibliography]{revtex4-1}

\usepackage{amssymb}
\usepackage{amsmath}
\usepackage{graphicx}
\usepackage{subfigure}
\usepackage{textcomp}
\usepackage{color}
\usepackage{amsfonts}
\usepackage{bbold}
\usepackage{dsfont}
\usepackage{epsfig}
\usepackage{hyperref}
\usepackage{array}

\begin{document}
\title{Computational and experimental investigation for new transition metal selenides and sulfides: The importance of experimental verification for stability}

\author{Awadhesh Narayan}
\affiliation{Department of Physics, University of Illinois at Urbana-Champaign, Urbana, Illinois, USA.}
\author{Ankita Bhutani}
\affiliation{Department of Materials Science and Engineering, University of Illinois at Urbana-Champaign, Urbana, Illinois, USA.}
\author{Samantha Rubeck}
\affiliation{Department of Physics, University of Illinois at Urbana-Champaign, Urbana, Illinois, USA.}
\author{James N. Eckstein}
\affiliation{Department of Physics, University of Illinois at Urbana-Champaign, Urbana, Illinois, USA.}
\author{Daniel P. Shoemaker}
\affiliation{Department of Materials Science and Engineering, University of Illinois at Urbana-Champaign, Urbana, Illinois, USA.}
\author{Lucas K. Wagner}
\affiliation{Department of Physics, University of Illinois at Urbana-Champaign, Urbana, Illinois, USA.}
\date{\today}

\begin{abstract}
Expanding the library of known inorganic materials with functional electronic or magnetic behavior is a longstanding goal in condensed matter physics and materials science.
Recently, the transition metal chalchogenides including selenium and sulfur have been of interest because of their correlated-electron properties, as seen in the iron based superconductors and the transition metal dichalcogenides. 
However, the chalcogenide chemical space is less explored than that of oxides, and there is an open question of whether there may be new materials heretofore undiscovered.
We perform a systematic combined theoretical and experimental search over ternary phase diagrams that are empty in the Inorganic Crystal Structure Database containing cations, transition metals, and one of selenium or sulfur.
In these 27 ternary systems, we use a probabilistic model to reduce the likelihood of false negative predictions, which results in a list of 24 candidate materials.
We then conduct a variety of synthesis experiments to check the candidate materials for stability.
While the prediction method did obtain previously unknown compositions that are predicted stable within density functional theory, none of the candidate materials formed in our experiments.
We come to the conclusion that these phase diagrams are  ``empty'' in the case of bulk synthesis, but it remains a possibility that alternate synthesis routes may produce some of these phases.
\end{abstract}

\maketitle

\section{Introduction}

In the last decade or so, high-throughput searches for new materials have been gaining increasing interest as novel and rapid routes to find stable compounds. 
Spurred by the increasing availability of large-scale computational resources as well as the rise of a number of materials databases, it has emerged as a rapidly evolving field of materials science.~\cite{curtarolo2013high} 
There already exist a number of computational materials repositories, including Materials Project,~\cite{jain2013commentary} {\sc aflow} library,~\cite{curtarolo2012aflowlib} Open Quantum Materials Database,~\cite{saal2013materials} Computational Materials Repository,~\cite{compmatrep} and the Electronic Structure Project,~\cite{elecstrucproj} to name a few. 
These databases offer  opportunities for design and discovery of new materials based on data-mining and machine-learning approaches.

In recent years this high-throughput machinery has been employed in a number of studies to search for potentially stable oxides. Using a combination of machine learning and first-principles calculations, Hautier and co-workers have explored the possibility of finding ``nature's missing ternary oxides.''~\cite{hautier2010finding} 
High-throughput \textit{ab initio} computations have also been used to evaluate another class of oxygen based compounds, namely phosphates, as Lithium-ion battery cathodes.~\cite{hautier2011phosphates} 
A systematic experimental investigation of high-throughput theoretical predictions and their structural characterization, which serves to validate the computational results, has also been performed for alkali carbonophosphate compounds.~\cite{chen2012synthesis} 
A computational search for $p$-type transparent conducting oxides has also been undertaken.~\cite{hautier2013identification} 
Very recently, Cu, Ag and Au ternary oxides have been identified using a high-throughput search based on the minima hopping method.~\cite{cerqueira2015identification}

In our searched chemical space of chosen cations and $3d$ transition metals, more than two thousand three hundred oxides are listed in Inorganic Crystal Structure Database (ICSD), whereas only around three hundred and fifty selenides and sulfides have been reported. While significant effort has been devoted to high-throughput searches for oxides, comparatively little is known about the missing selenides and sulfides, and whether they might exist in as-yet empty ternary phases. In this contribution we take a first step along this direction for a number of ternary $3d$ transition metal selenides and sulfides, by identifying, \textit{in silico}, which of these can be stable. 
Using a data-mining based ionic substitution model, in conjunction with density functional theory computations we study ternary systems of the form XYZ, where the cation X= Ba, Ca, Sr, La, K, Bi, Pb, Y is a $3d$ transition metal and Z= S or Se. There are 27 empty ternary phase diagrams with such a combination of elements. 
Based on a Bayesian likelihood estimate to minimize false negative predictions, we narrow down the search to 24 candidates.
Three of these candidates are very similar to misfit structures not present in the ICSD. Our predicted structures are similar, though not identical, to these disordered structures and within the framework of the ionic substitution model we only consider ordered structure types.
We performed high-throughput experiments to synthesize the remaining candidates, using solid state and gas flow reactions. 
Our experimental attempts resulted in only binary compounds for these systems. 
Based on this combined theoretical and experimental study we conclude that the ternary compounds in these 27 phase diagrams, are not accessible via standard bulk synthesis methods. Nevertheless, it is important for us to present their predicted proximity to stability, and document our experimental attempts, since other synthesis methods may produce these phases.

\section{Computational Methods}\label{methods}

\subsection{Generation of candidate structures}
We generated structures using the method of Hautier \textit{et al.}~\cite{hautier2010data}, in which ions are substituted into existing known crystal structures to form new compounds.
The probability for such substitutions is obtained by constructing feature functions, which capture the key aspects of these replacements. 
Feature functions are augmented by weights for substitutions, which are extracted from ternary and quaternary ionic compounds in the ICSD.~\cite{icsd} 
These are used to evaluate the probabilities and assign ranks to the generated structures.
In this way, the ionic substitution method can generate likely structural candidates very quickly. 

In the following we briefly summarize the main features of this model and refer the readers to Ref.~\onlinecite{hautier2010data} for a more detailed exposition. The mathematical formulation of this model begins by representing any compound composed of ions, $X_{i}$, ($i=1,...,n$) as an $n$ component vector
\begin{equation}
 \mathbf{X}=(X_{1},X_{2},...,X_{n}).
\end{equation}
One then requires the probability, $p_{n}$ for two different compounds having the same number of components to exist in the same crystal structure. This is expressed as
\begin{equation}
 p_{n}(\mathbf{X},\mathbf{X'})=p_{n}(X_{1},...,X_{n},X_{1}',...,X_{n}').
\end{equation}
This probability function needs to be approximated. In the ionic substitution model this is achieved by using feature functions, $f(\mathbf{X},\mathbf{X'})$, which capture the key aspects of the problem. A mathematical requirement for the feature function is that it should be defined on the domain of the sought probability function and should return either 0 or 1. One can then approximate the probability function as
\begin{equation}
 p_{n}(\mathbf{X},\mathbf{X'})\approx \frac{\exp\left[{\sum_{i}\lambda_{i}f_{i}(\mathbf{X},\mathbf{X'})}\right]}{Z},
\end{equation}
 where $\lambda_{i}$ is the weight corresponding to feature $f_{i}(\mathbf{X},\mathbf{X'})$, and $Z$ is analogous to a partition function which ensures normalization of the probability function. An implicit assumption is that the feature functions do not depend on the number of ions, $n$, in the compound, i.e., ionic substitution rules do not depend on $n$. Then, one assumes that the probability of two ions to replace each other is the same regardless of other ions in the compound. This is implemented by considering only binary feature functions, which read
\begin{eqnarray}
 f_{k}^{\alpha,\beta}(\mathbf{X},\mathbf{X'}) &=& 1 \quad X_{k}=\alpha, X_{k}'=\beta \nonumber \\
 &=& 0 \quad \mathrm{otherwise}
\end{eqnarray}

 Every pair of ions $(\alpha,\beta)$ is assigned feature functions of the above form. 
 The corresponding weights, $\lambda_{k}^{\alpha,\beta}$, indicate how likely is the $\alpha\leftrightarrow\beta$ binary substitution. These weights are obtained from the ternary and quaternary ionic compounds in the Inorganic Crystal Structure Database (ICSD).~\cite{icsd} From this database, crystal similarities are then estimated using structure comparison algorithms.~\cite{hundt2006cmpz} Then, one can assign specific values $(\mathbf{x},\mathbf{x'})$ to the vectors $(\mathbf{X},\mathbf{X'})$. For the full database this leads to 
\begin{equation}
 \mathbf{D}=\left\{(\mathbf{X},\mathbf{X'})=(\mathbf{x},\mathbf{x'})^{1},...,(\mathbf{x},\mathbf{x'})^{m}\right\}.
\end{equation}
 Using the above assignments, the maximum-likelihood approach is then employed to estimate the vector of weights $\mathbf{\lambda}$. The log-likelihood $l$ of the observed data is
\begin{equation}
l(\mathbf{D},\mathbf{\lambda})=\sum_{j=1}^{m}\left[\sum_{i}\lambda_{i}f_{i}((\mathbf{x},\mathbf{x'})^{j})-\log Z(\mathbf{\lambda})\right].
\end{equation}
 The weights which maximize this log-likelihood, $l(\mathbf{D},\mathbf{\lambda})$, are chosen as the weights for the probabilistic ionic substitution model.

The ionic substitution method is particularly attractive for the problem we are considering, since we are interested in a survey over 27 previously empty phase diagrams.
Techniques that directly search for the minimum in the free energy landscape such as Refs.~\onlinecite{oganov2006crystal,trimarchi2007global}, while they may be preferable in terms of calculating the overall lowest energy structures, would be computationally prohibitive for this study. 

Once we generated the candidate structures, we followed a multi-level strategy to assess the stability of these structures. 
The first level of filtering is obtained by choosing the one hundred candidate structures with highest probabilities assigned by the ionic substitution model. 
At the next step we performed density functional theory (DFT) calculations for these selected candidates using Quantum Espresso package.~\cite{giannozzi2009quantum} 
We used the Perdew-Burke-Ernzerhof (PBE) parameterization to the exchange-correlation functional.~\cite{perdew1996generalized} 
To replace the core electrons we used Rappe-Rabe-Kaxiras-Joannopoulos relativistic ultrasoft pseudopotentials.~\cite{rappe1990optimized} All the candidate structures were relaxed until the forces on atoms were less than 0.01 eV/\AA{} and the pressure was less than 0.5 kbar. Other computational parameters were chosen following the materials project, which have been shown to lead to total energies being converged to less than 5~meV/atom.~\cite{jain2011high} 
For all calculations, we started with a ferromagnetic initialization of spins. 
Finally, we constructed the phase diagrams by comparing the total energies for the candidate compounds with all known binaries and elements making up the compound. The total energies of elements and binaries were calculated with our calculation setup consistent with that for the ternaries.
We used the tools available in Python Materials Genomics ({\sc pymatgen}) package to analyze the data.~\cite{ong2013python} 

\subsection{Decisions based on DFT data}

Structural phase diagrams require a high degree of accuracy, which DFT calculations do not attain. 
For example, according to Jain \textit{et al.}\cite{jain2013commentary}, the DFT-calculated hull energies in the searched chemical space are close to normally distributed. Only 40\% of the 106 existing ternaries in the searched chemical space lie on the hull. 
Ultimately, we wish to make a binary decision: we either attempt to synthesize a compound or not, based on the DFT calculations. 
There are thus two potential errors one can make: false positive, in which a compound is suggested for synthesis and does not form, and false negative, in which a candidate compound is discarded but would have formed.

\begin{figure*}[t]
\begin{center}
  \includegraphics{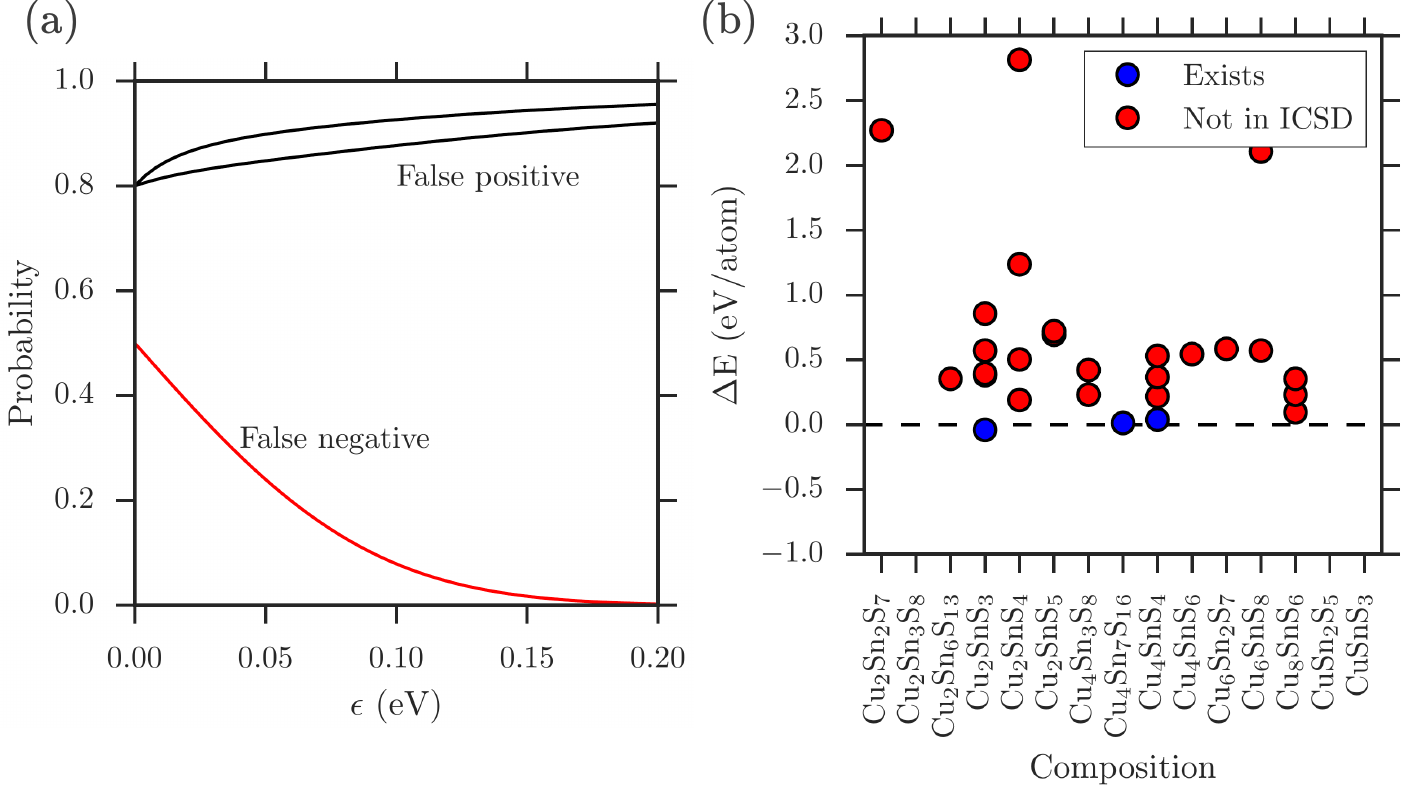}
  \caption{(Color online) (a) Estimates of false positive and negative rates as a function of the cutoff energy $\epsilon$ for stability prediction using DFT-PBE. Two false positive curves are presented, for $\alpha$=100 and 200 structures/eV. (b) Hull energies for candidate structures in the Sn-Cu-S ternary system. The known structures listed in ICSD are shown as blue circles. Structures with hull energies greater than 3 eV/atom are not shown.} 
  \label{fig:probability_model}
\end{center}
\end{figure*}

We wish to consider the probability that a single structure is experimentally stable, given that the DFT-calculated enthalpy of formation is $\Delta H$ relative to the known and proposed materials in the phase diagram.
In this notation, we include all known compounds in the phase diagram, so that if $\Delta H$ is negative, we expect the candidate compound to be stable.
Using Bayes' theorem, we can write
\begin{equation}
P(S|\Delta H < \epsilon) = \frac{P(\Delta H < \epsilon|S)P(S)}{P(\Delta H < \epsilon)}
\end{equation}
where $\epsilon$ is an enthalpy of formation cutoff that we will choose and $S$ means that the candidate compound is stable.

We can estimate the quantities as follows:
\begin{itemize}
\item 	$\frac{P(S)}{P(\Delta H < \epsilon)}$ : In a known phase diagram, the number of stable ternaries divided by the number of ternaries with $\Delta H < \epsilon$
\item $P(\Delta H < \epsilon|S)$ : For known stable ternaries, the number of materials with $\Delta H < \epsilon$ divided by the number of ternaries considered. This must be evaluated for all structures generated by the ionic substitution model.
\end{itemize}
To estimate the probabilities, let us assume a Gaussian distribution for $P(\Delta H < \epsilon|S)$ with a standard deviation of 0.1 eV. 
We estimate $\frac{P(S)}{P(\Delta H < \epsilon)}\simeq \frac{N_{S}}{N_{0}+\alpha\epsilon}$. 
$N_{S}$ is the number of experimentally stable structures, while $N_{0}$ is the number of proposed structures with $\Delta H < 0$. 
Here the proposed structures are the ones screened by the ionic substitution model. 
The estimates for $N_{0}$ and $\alpha$ are obtained from ternary phase diagrams in the same chemical space which have known ternary compounds. 
We used the ionic substitution model and total energy calculations to estimate $N_0\simeq 5$ and $\alpha \simeq 100$ structure/eV. 
$N_0$ controls the y-intercept of Fig.~\ref{fig:probability_model}, and $\alpha$ partially affects the slope. 
Since $N_0$ and $\alpha$ are the most uncertain values in this model, we have checked that the results are reasonably robust to their choice as shown in Fig.~\ref{fig:probability_model}.
To demonstrate that our choices are reasonable, we display hull energies for Sn-Cu-S system from the Materials Project\cite{jain2013commentary} in Fig.~\ref{fig:probability_model}(b). 
Our parameters also agree with extensive cross validation for known phase diagrams in the literature, such as that undertaken by Hautier and coauthors~\cite{hautier2010data}.

When choosing $\epsilon$, we need to consider both the false positive rate:
\begin{equation}
P(NS|\Delta H < \epsilon) = \frac{P(\Delta H < \epsilon|NS) P(NS)}{P(\Delta H < \epsilon)}= 1- P(S|\Delta H < \epsilon),
\end{equation}
and the false negative rate:
\begin{equation}
P(\Delta H > \epsilon|S) = \left(1-P(\Delta H < \epsilon|S)\right)
\end{equation}
Estimated probabilities of false positive and false negative identifications are shown in Fig.~\ref{fig:probability_model}, with varying cutoff, $\epsilon$, and a choice of $N_{S}=2$ and $N_{0}=5$. 
To avoid missing new compounds, we would like to minimize the likelihood of false negative predictions and we set $\epsilon=0.1$~eV. 
However, this means that our false positive rate ends up being quite large, or in other words the true positive rate would be low. 
For our choice of $\epsilon=0.1$~eV, the probability of true positive predictions is around 0.05-0.1, depending on the values of $\alpha$ and $N_0$. 
Taking into account the errors associated with the prediction method our model allows us to estimate the false positive and false negative rates associated with our predictions. 
As we will report later in the paper, our high-throughput experiments find 3 out of the predicted 24 candidates to be stable. 
This ratio is close to our Bayesian estimate of true positive predictions. 
This good agreement suggests that our Bayesian model provides a reasonable framework to analyze the reliability of the prediction methods. 
It also provides insight that the inherent uncertainties in DFT energies leads to a high false positive rate, if the cutoff is set so that the false negative rate is low. 
It also sheds light on the reason behind why many of the DFT-predicted compounds were not synthesized in our experiments.

\section{Computational results}\label{results}

\begin{table*}[t]
\caption{Summary of candidate compounds found in this work: Chemical formula, energy above hull ($E_{AH}$) or inverse hull energy ($E_{IH}$, in (eV/atom)), magnetic moment ($\mu$, in $\mu_{B}$/TM), DFT-PBE band gap (in eV), structure type, transition metal-anion geometry and experimental status for candidate compounds. Transition metal complexes form tetrahedra which are edge sharing (edge-tet), vertex sharing (vertex-tet), both edge and vertex sharing (edge-vertex-tet) or isolated (isolated-tet). Six-fold coordination was limited to edge sharing octahedra (edge-oct). }
\centering 
\begin{tabular}{c c c c c c c} 
\hline \hline 
Formula & $E_{AH}$/$E_{IH}$ & $\mu$ & Gap & Structure type & Geometry & Forms?\\[0.5ex]
\hline \hline
SrMn$_{2}$S$_{4}$ & -0.028 & 1.74 & 0.00 & PbC$_2$O$_4$ & edge-tet & No\\
CaMn$_{2}$S$_{4}$ & +0.068 & 1.76 & 0.00 & PbC$_2$O$_4$ & edge-tet & No\\
SrFe$_2$Se$_{3}$ & -0.095 & 2.73 & 0.00 & BaFe$_2$Se$_3$ & edge-tet& No\\
PbFe$_2$Se$_{3}$ & -0.068 & 2.51 & 0.00 & BaFe$_2$Se$_3$ & edge-tet& No (MBE)\\
Pb$_2$FeSe$_{3}$ & -0.095 & 3.67 & 0.00 & Ba$_2$CdSe$_3$ & edge-tet& No (MBE)\\
SrNiS$_{2}$ & +0.068 & 0.00 & 0.00 & BaNiS$_2$ & edge-tet & No \\
SrMn$_{4}$S$_{7}$ & +0.027 & 3.50 & 0.00 & SrB$_{4}$O$_{7}$ & vertex-tet & No\\
Sr$_2$FeSe$_{3}$ & -0.094 & 3.23 & 0.00 & Eu$_2$CuS$_3$ & vertex-tet& No (MBE)\\
Ba$_{2}$Sc$_{2}$S$_{5}$ & +0.068 & 0.00 & 2.61 & Sr$_2$Ga$_2$S$_5$ & vertex-tet & -- (Sc)\\
SrCu$_{2}$Se$_{2}$ & +0.054 & 0.00 & 0.32 & BaCu$_2$S$_2$ & edge-vertex-tet& -- (Cu$^{1+}$)\\
SrCu$_{2}$S$_{2}$ & +0.027 & 0.00 & 0.55 & BaCu$_2$S$_2$ & edge-vertex-tet & -- (Cu$^{1+}$)\\
LaFeSe$_2$ & -0.176 & 2.81 & 0.00 & CuLaS$_2$ & edge-vertex-tet & No\\
Sr$_{3}$FeSe$_{5}$ & -0.081 & 0.16 & 0.26 & SrBeEu$_2$O$_5$ & isolated-tet& No\\
Pb$_2$FeSe$_{4}$ & -0.014 & 0.01 & 0.19 & Pb$_2$SiSe$_4$ & isolated-tet & No (MBE)\\
Sr$_{2}$VS$_{4}$ & -0.013 & 0.92 & 0.00 & K$_2$WSe$_4$ & isolated-tet & No \\
K$_{6}$ScS$_{4}$ & +0.014 & 0.01 & 0.00 & K$_{6}$HgS$_{4}$ & isolated-tet & -- (Sc)\\
BiFeSe$_{3}$ & -0.095 & 1.08 & 0.52 & SbCrSe$_3$ & edge-oct & No\cite{ranmohotti_coexistence_2015}\\
La$_{3}$ScS$_{6}$ & +0.013 & 0.00 & 0.97 & La$_3$InS$_6$ & edge-oct & -- (Sc)\\
SrV$_{4}$S$_{7}$ & -0.014 & 1.51 & 0.00 & BaCr$_4$S$_7$ & edge-oct & No \\
SrCr$_{4}$S$_{7}$ & +0.028 & 2.74 & 0.00 & BaCr$_4$S$_7$ & edge-oct& No \\
KScS$_{2}$ & -0.136 & 0.00 & 1.67 & CsPrS$_2$ & edge-oct & -- (Sc)\\
LaTiS$_{3}$ & +0.027 & 0.02 & 0.00 & NH$_4$CdCl$_3$ & edge-oct& Yes(misfit)\\
LaVS$_{3}$ & -0.028 & 1.92 & 0.00 & NH$_4$CdCl$_3$ & edge-oct& Yes(misfit)\\
LaMnS$_{3}$ & -0.095 & 3.47 & 0.00 & NH$_4$CdCl$_3$ & edge-oct & Yes(misfit)\\
\hline \hline
\end{tabular}\label{table_summary}
\end{table*}

\begin{figure*}[t]
\begin{center}
  \includegraphics[scale=1.0]{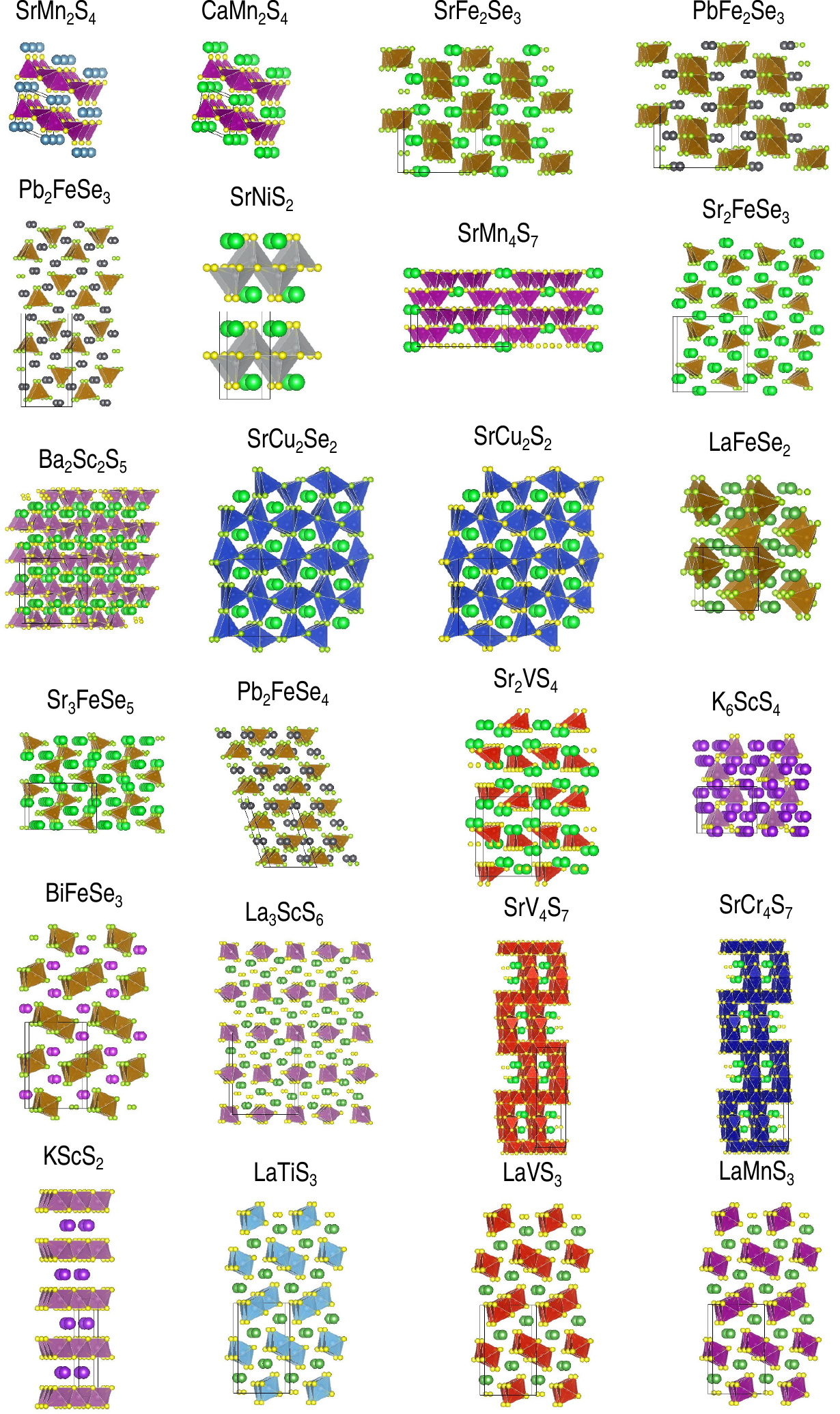}
  \caption{(Color online) Structures of the candidate compounds. The transition metal is depicted enclosed in a polyhedron formed by the small yellow/green spheres which represent S/Se atoms, while the third larger sphere depicts the other cation and black lines denote the unit cell for the crystal.} \label{collected_structures}
\end{center}
\end{figure*}

Our results for the candidate compounds are summarized in Table~\ref{table_summary} and their structures are shown in Fig.~\ref{collected_structures}. We find that in nearly two-thirds of our predicted compounds, the transition metal is in four-fold coordination. These tetrahedra can be either edge sharing, vertex sharing, both edge as well as vertex sharing or can be isolated from each other. In the remaining compounds the transition metal-anion octahedra are edge sharing. We also observe that a number of compounds form one-dimensional chains. 
Here we would like to make some general comparisons  of our predicted class of compounds with the existing ternary systems within the same chemical space. 
A majority of compounds that we have predicted have metallic character according to DFT-PBE (sixteen out of a total of twenty four). 
This ratio is similar to that of the known ternaries with the same constituent elements, where nearly sixty compounds (out of the existing 106 compounds) have a zero band gap. 
However, PBE is known to underestimate band gaps, and these values should be taken with caution. 
Among our predicted compounds, fifteen have a magnetic moment greater than 0.05 $\mu_{B}$ per transition metal atom. 
In the existing ternaries the ratio of ones with magnetic moment is very similar, with nearly fifty five percent of the compounds showing a finite magnetic moment. 
Crystallographic information files for all compounds is provided in the supplemental material.~\cite{supplement} 

\begin{figure}[t]
\begin{center}
  \includegraphics[scale=1.0]{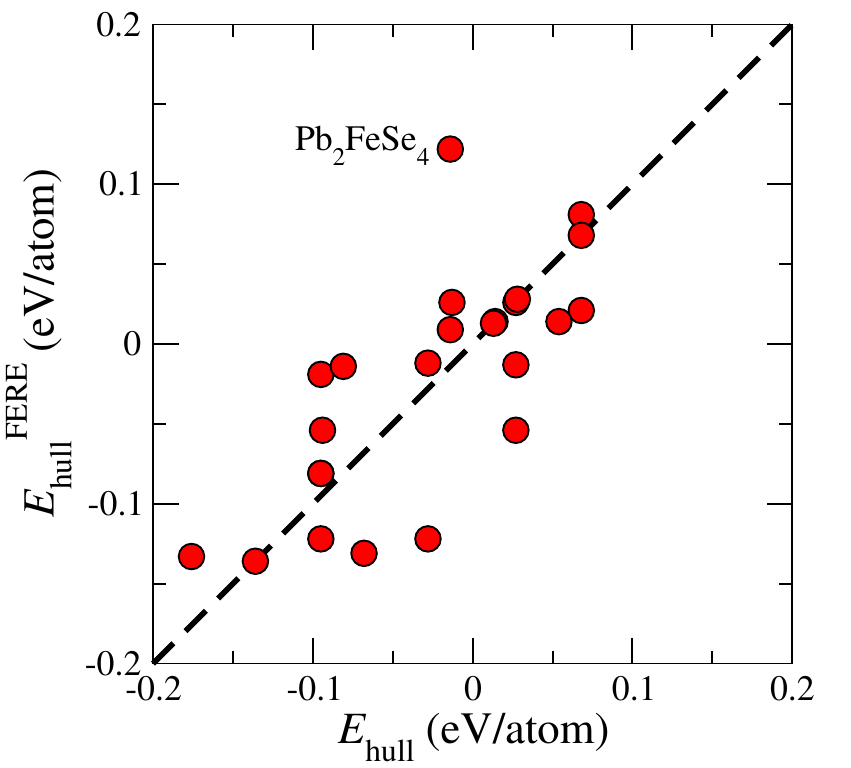}
  \caption{(Color online) Comparison of hull energies obtained without and with the fitted elemental-phase reference energy correction.} 
  \label{fig:hull_energy_fere}
\end{center}
\end{figure}

We have also computed the enthalpies of formation of our candidate compounds including a correction which adjusts the energies of the elements to correct for the errors known to affect the values predicted by DFT. 
We use the fitted elemental-phase reference energy (FERE) correction proposed by Stevanovic \textit{et al}.~\cite{stevanovic2012correcting} 
Following these authors, we used the PBE+U method, using $U=5.0$ for Cu and $U=3.0$ for all other $3d$ transition metals. 
A comparison of the hull energies without and with the FERE correction is shown in Fig.~\ref{fig:hull_energy_fere}. 
For most candidate compounds, the hull energies calculated without and with the correction are close enough to one another that we would not have changed our decision making process. 
A pronounced discrepancy appears for Pb$_2$FeSe$_{4}$ because the FERE correction for Pb has not been reported and we, therefore, did not apply the correction for Pb.

\begin{figure*}[t]
\begin{center}
  \includegraphics[scale=1.0]{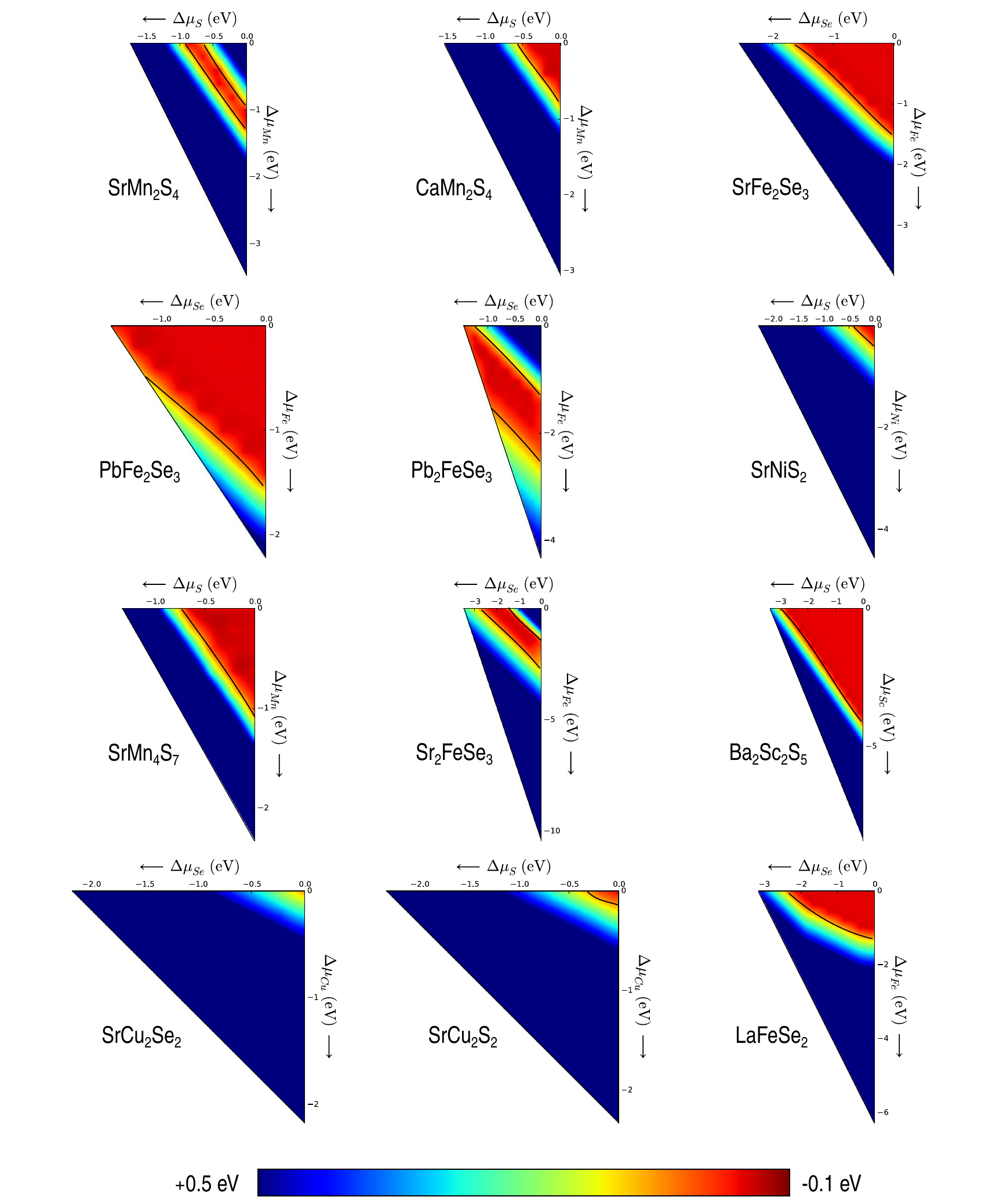}
  \caption{(Color online) Heat maps of hull energies with varying anion and transition metal chemical potentials. Black lines mark the chemical potential region where the candidate compound is stable.} 
  \label{fig:chem_pot1}
\end{center}
\end{figure*}

\begin{figure*}[t]
\begin{center}
  \includegraphics[scale=1.0]{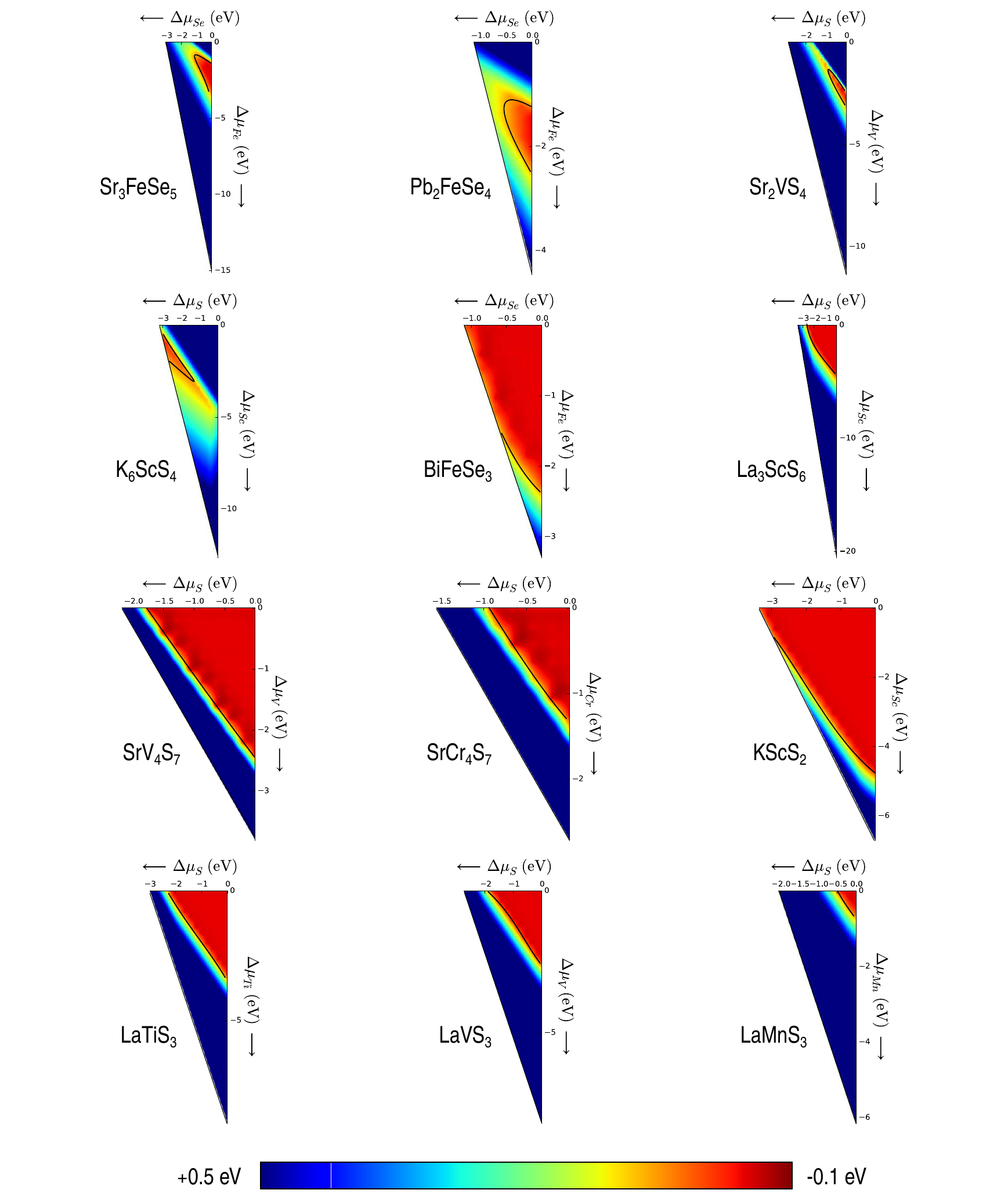}
  \caption{(Color online) Heat maps of hull energies with varying anion and transition metal chemical potentials. Black lines mark the chemical potential region where the candidate compound is stable.} 
  \label{fig:chem_pot2}
\end{center}
\end{figure*}

Varying the chemical potentials of different constituent elements provides an additional control knob and allows one to stabilize different compositions in the phase diagram.~\cite{persson2005n} 
The variation of hull energies for our candidate compounds, as a function of chemical potential of the anion and the transition metal are shown in Figs.~\ref{fig:chem_pot1} and \ref{fig:chem_pot2}. 
The optimal range for synthesis is marked by the black lines and may provide guidance to future attempts to synthesize these materials. 


Let us next categorize the compounds based on their structural motifs and discuss their electronic and magnetic properties in more detail.

\subsection{Compounds with edge-sharing tetrahedra}

\begin{figure*}[t]
\begin{center}
  \includegraphics[scale=1.0]{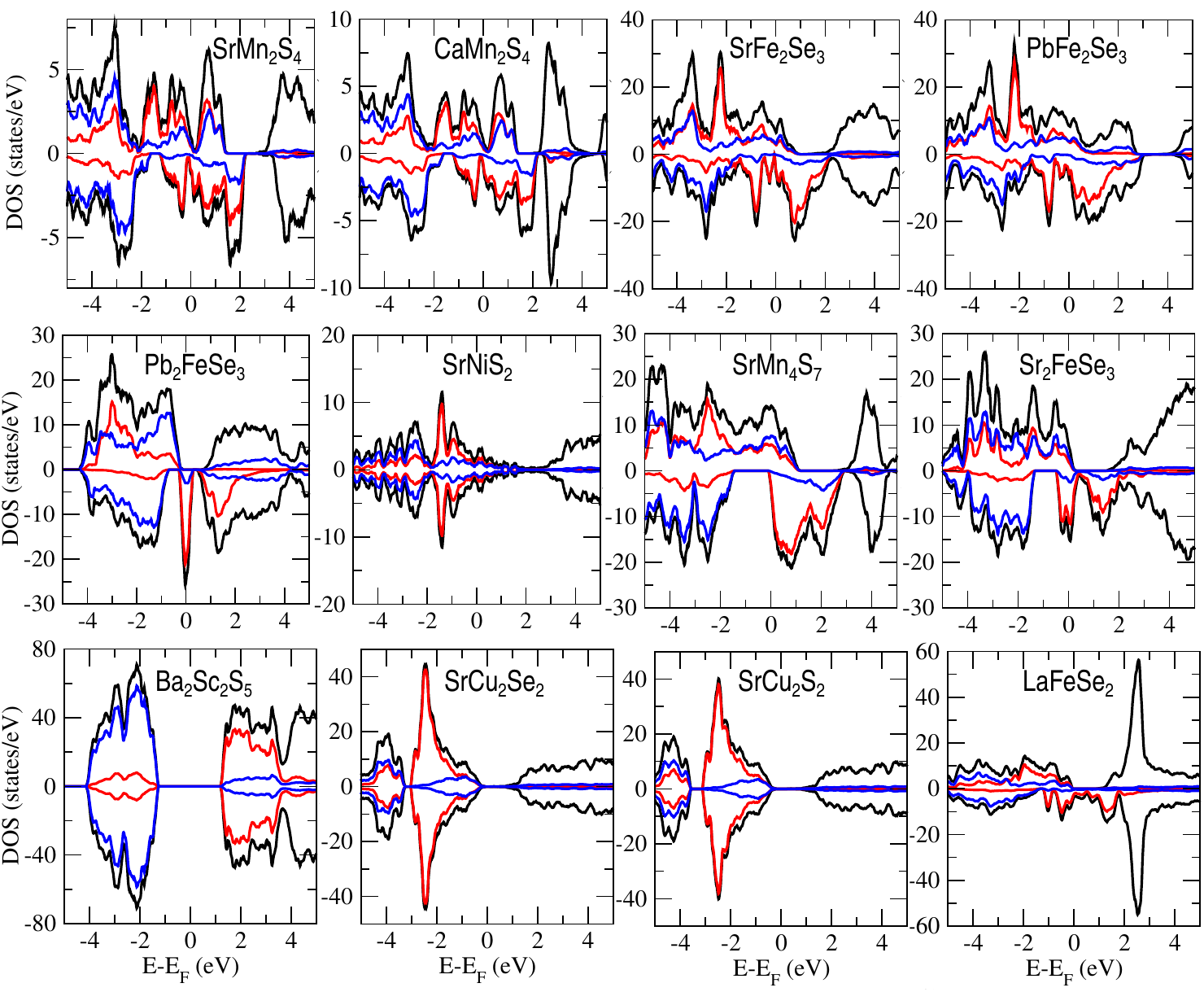}
  \caption{(Color online) Densities of states of the candidate compounds. Here black curves are the total density of states, while red and blue curves are the density of states projected on to transition metal (TM) $d$ orbitals and anion $p$ orbitals, respectively.} 
  \label{collected_dos_set1}
\end{center}
\end{figure*}

We begin with the class of ternaries comprising of transition metals and anions forming edge-sharing tetrahedra. Two of these,  SrMn$_{2}$S$_{4}$ and CaMn$_{2}$S$_{4}$, are isostructural, crystallizing in the PbC$_2$O$_4$ parent structure type. As shown in Fig.~\ref{collected_structures}, Mn atoms are coordinated in edge sharing tetrahedra forming one-dimensional chains. While CaMn$_{2}$S$_{4}$ lies slightly above the convex hull ($E_{AH}=0.068$ eV/atom), SrMn$_{2}$S$_{4}$ is located on the convex hull, with an inverse hull energy, $E_{IH}=-0.028$ eV/atom. Both these materials show a very similar electronic structure, with Sr to Ca replacement having negligible effects. This can be seen from the density of states (DOS) for the two compounds [compare top left panels in Fig.~\ref{collected_dos_set1}]. Mn $d$ states in the two compounds are partially filled, resulting in a small magnetic moment.

The next two materials in this category, SrFe$_2$Se$_{3}$ and PbFe$_2$Se$_{3}$ are formed in BaFe$_2$Se$_3$ parent structure type. The structure consists of one-dimensional chains with the transition metal atoms forming edge sharing tetrahedra [Fig.~\ref{collected_structures}]. SrFe$_2$Se$_{3}$ and PbFe$_2$Se$_{3}$ have reasonably large inverse hull energies of -0.095 eV/atom and -0.068 eV/atom, respectively. Similar to the case of SrMn$_{2}$S$_{4}$ and CaMn$_{2}$S$_{4}$, SrFe$_2$Se$_{3}$ and PbFe$_2$Se$_{3}$ show nearly identical electronic structure, with cation replacement from Sr to Pb leaving the DOS largely unchanged. For both these materials the DOS at the Fermi level ($E_{F}$) comes from an almost equal admixture from Fe $d$ levels and Se $p$ states, as shown in top right panels of Fig.~\ref{collected_dos_set1}. 

Pb$_2$FeSe$_{3}$ has an inverse hull energy of -0.095 eV/atom and forms in Ba$_2$CdSe$_3$ structure type as shown in Fig.~\ref{collected_structures}. For Pb$_2$FeSe$_{3}$, within DFT-PBE, there is a finite DOS at the Fermi level for the down spin electrons, while the up spin DOS is zero. DOS projected (PDOS) onto transition metal $d$ orbitals and anion $p$ orbitals indicate that the DOS peak at $E_{F}$ arises primarily from the Fe $d$ states. SrNiS$_{2}$ lies slightly above the convex hull and forms a layered structure [Fig.~\ref{collected_structures}]. The compound is a metal with an almost equal contribution to the DOS near the Fermi level coming from Ni $d$ states and S $p$ levels, as shown in Fig.~\ref{collected_dos_set1}. Ni occupied $d$ states lie around 1 eV below the Fermi level, and the compound has zero magnetic moment.

\subsection{Compounds with corner-sharing tetrahedra}

\begin{figure*}[t]
\begin{center}
  \includegraphics[scale=1.0]{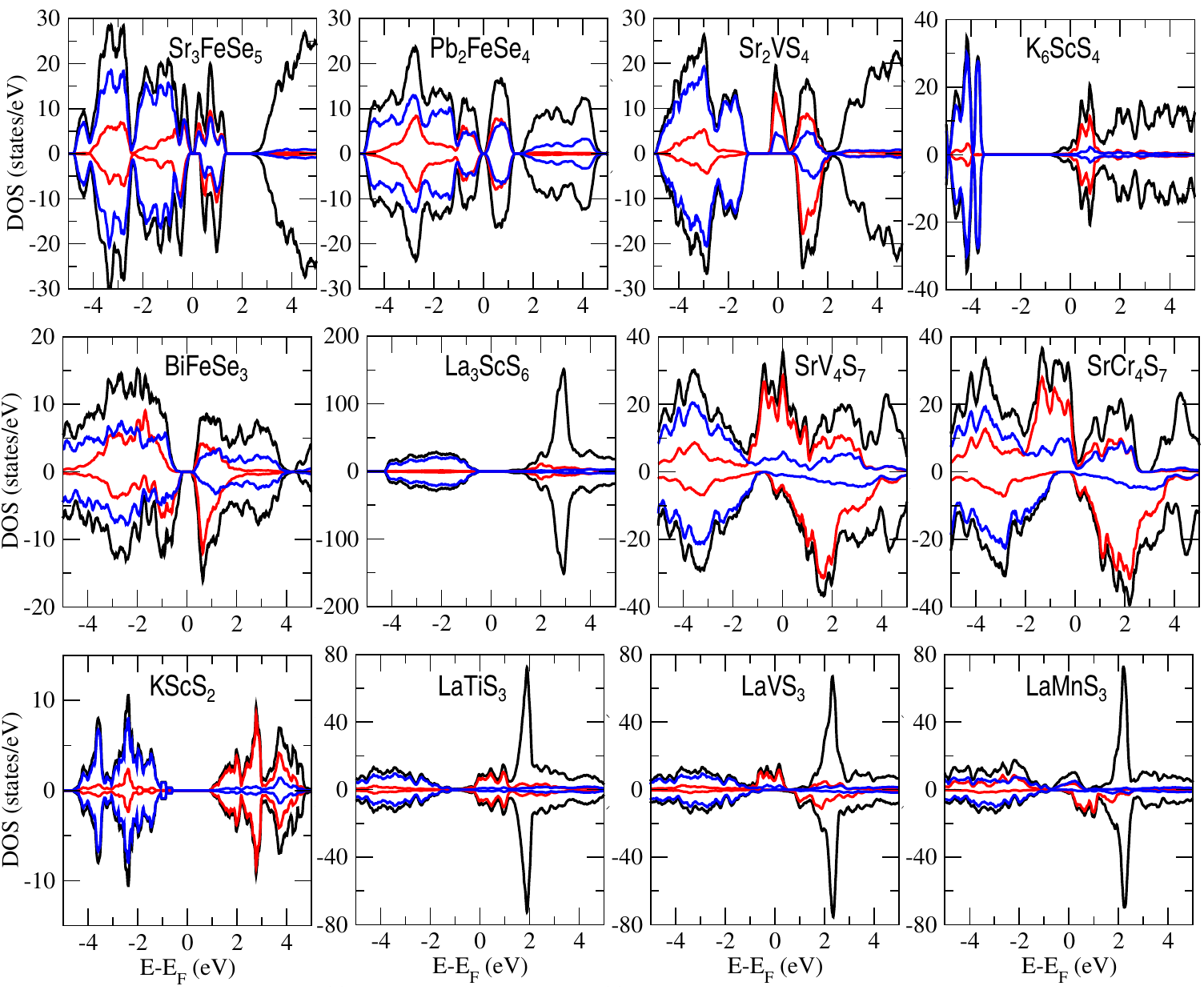}
  \caption{(Color online) Densities of states of the candidate compounds. Here black curves are the total density of states, while red and blue curves are the density of states projected on to transition metal (TM) $d$ orbitals and anion $p$ orbitals, respectively.} 
  \label{collected_dos_set2}
\end{center}
\end{figure*}

We move on to the compounds where the transition metal forms corner-sharing tetrahedra with S or Se atoms. In this category, SrMn$_{4}$S$_{7}$ is marginally unstable, being 0.027 eV/atom above the convex hull. It forms a layered structure, with each S atom being shared between three Mn tetrahedra (SrB$_{4}$O$_{7}$ structure type) as illustrated in Fig.~\ref{collected_structures}. The densities of states for the compound is shown in the middle panel of Fig.~\ref{collected_dos_set1}, indicating the contribution from both spins at the Fermi level. PDOS plots reveal, however, that near the Fermi level the overlap between the transition metal $d$ states and anion $p$ levels occurs predominantly for the up spin, while being largely suppressed for the opposite spin. The next compound in this class, Sr$_2$FeSe$_{3}$, crystallizes in the Eu$_2$CuS$_3$ parent structure type, with Fe atoms forming vertex sharing tetrahedral chains, and has an inverse hull energy of -0.094 eV/atom. Ba$_{2}$Sc$_{2}$S$_{5}$ lies 0.068 eV/atom above the convex hull and crystallizes in the Sr$_2$Ga$_2$S$_5$ structure type, with Sc atoms being present in a tetrahedral coordination. It has a zero magnetic moment along with a large DFT-PBE band gap of 2.61 eV [Fig.~\ref{collected_dos_set1}]. The majority of the DOS contribution in the occupied states arises from S $p$ orbitals, with the Sc $d$ orbitals being predominantly empty.

\subsection{Compounds with both edge- and corner-sharing tetrahedra}

In the class of compounds with both edge- as well as corner-sharing tetrahedra, isostructural compounds SrCu$_{2}$Se$_{2}$ and SrCu$_{2}$S$_{2}$ are marginally above the hull by 0.054 eV/atom and 0.027 eV/atom, respectively. They are predicted to crystallize in the BaCu$_2$S$_2$ structure type. Cu atoms are in a tetrahedral coordination, with the tetrahedra lying in an alternating tilting arrangement, as shown in Fig.~\ref{collected_structures}. We find that both the materials show a finite DFT-PBE band gap, as can be seen from the bottom center panels of Fig.~\ref{collected_dos_set1}. Filled Cu $d$ orbitals in both these compounds lie nearly 3 eV below the Fermi level, and the magnetic moments are zero. LaFeSe$_2$ has a substantial inverse hull energy of -0.176 eV/atom and is the most stable compound that we have found. It crystallizes in CuLaS$_2$ parent structure. As shown in Fig.~\ref{collected_dos_set1}, LaFeSe$_2$ has a peculiar density of states, where there is a small, but finite, DOS at the Fermi level for the down spin electrons, while the up spin DOS is negligible. LaFeSe$_2$ also has a large magentic moment of 2.81 $\mu_{B}$/Fe atom.

\subsection{Compounds with isolated tetrahedra}

We next discuss the compounds in which the transition metal atom is in a four-fold coordination with the anion, and these tetrahedra are isolated from each other. The first two materials in this class, Sr$_{3}$FeSe$_{5}$ and Pb$_2$FeSe$_{4}$, are predicted to crystallize in SrBeEu$_2$O$_5$ and Pb$_2$SiSe$_4$ structure types, respectively. We find that these two compounds have small DFT-PBE band gaps of around 0.2 eV, as seen from the densities of states shown in Fig.~\ref{collected_dos_set2}. Close to the edges of the band gap energy window, there is a substantial overlap of Fe $d$ states and Se $p$ states. 

The next compound, Sr$_{2}$VS$_{4}$, has a small inverse hull energy of -0.013 eV/atom and it forms alternating tilted isolated tetrahedra, as shown in Fig.~\ref{collected_structures}. For Sr$_{2}$VS$_{4}$, there is a majority spin peak in the DOS at the Fermi level, while the down spin DOS vanishes. The main contribution to this peak arises from the $d$ orbitals of V atoms and the compound has a magnetic moment of nearly 1 $\mu_{B}$/V atom. The last compound in this category, K$_{6}$ScS$_{4}$, lies marginally above the hull by 0.014 eV/atom. Its crystal structure comprises of Sc-S tetrahedra surrounded by cages of K atoms. Expectedly, the compound is non-magnetic and the $d$ states of Sc are unoccupied.

\subsection{Compounds with edge sharing octahedra}

We next discuss the compounds where transition metal forms edge sharing octahedra with S or Se atoms. The first compound in this category, BiFeSe$_{3}$, crystallizes in SbCrSe$_3$ structure type, with the Fe atoms in one-dimensional chains formed by edge sharing octahedra. BiFeSe$_{3}$ has a 0.52 eV DFT-PBE gap, with nearly equal contribution to the density of states from Fe $d$ levels and Se $p$ states both in occupied and unoccupied manifold, around $E_{F}$ [Fig.~\ref{collected_dos_set2}]. Although, no ternaries in Bi-Fe-Se system are reported in the ICSD, very recently Bi$_2$FeSe$_4$ has been synthesized.~\cite{ranmohotti_coexistence_2015} La$_{3}$ScS$_{6}$, lies marginally above the convex hull ($E_{AH}=0.013$ eV/atom). It forms an alternating arrangement of Sc atoms in tetrahedral and octahedral coordination with S atoms. We find that the DFT-PBE band gap is nearly 1 eV, and similar to Ba$_{2}$Sc$_{2}$S$_{5}$, the Sc $d$ levels are unoccupied and most of the DOS contribution in the occupied manifold comes from S $p$ states. 

SrV$_{4}$S$_{7}$ and SrCr$_{4}$S$_{7}$ are predicted to crystallize in BaCr$_4$S$_7$ parent structure type. We have obtained both these compounds to be close to the convex hull, with SrV$_{4}$S$_{7}$ having an inverse hull energy of -0.014 eV/atom and SrCr$_{4}$S$_{7}$ lying +0.028 eV/atom above the hull. We find that these two materials have an interesting layered structure and the layers themselves are interconnected by one-dimensional chains, as shown in Fig.~\ref{collected_structures}. We can see from the density of states that SrV$_{4}$S$_{7}$ and SrCr$_{4}$S$_{7}$ have finite overlap of transition metal $d$ states and S $p$ states near the Fermi level for up spin, while the overlap is much reduced for the down spin. Furthermore, we find that in moving from V to the Cr compound, filling of the up spin $d$ states is enhanced, and as a result the magnetic moment increases from 1.51 $\mu_{B}$/V to 2.74 $\mu_{B}$/Cr. Finally, the last compound in this category, KScS$_{2}$, has a reasonably large inverse hull energy of -0.136 eV/atom. It has a layered structure, with K atoms forming a spacer layer separating the Sc-S layers [Fig.~\ref{collected_structures}]. We find that this material has a relatively large DFT-PBE band gap of 1.67 eV and no net magnetic moment.

We have found LaTiS$_{3}$, LaVS$_{3}$ and LaMnS$_{3}$ to be most stable in the NH$_4$CdCl$_3$ parent structure type. LaTiS$_{3}$ is obtained to be slightly above the convex hull, by 0.027 eV/atom, while both LaVS$_{3}$ and LaMnS$_{3}$ are found to lie on the convex hull, with inverse hull energies of -0.028 and -0.095 eV/atom, respectively. The crystal structures are shown in Fig.~\ref{collected_structures}. We obtain the three compounds to be metals within PBE. From the DOS projected onto the transition metal $d$ orbitals and the S anion $p$ orbitals, we see that they have a significant $d$ orbital contribution near the Fermi level. Expectedly, the magnetic moment increases as one changes the transition metal going from left to right in the periodic table, i.e., from Ti to V to Mn. We can also infer this from the PDOS of transition metal $d$ states. As we go from Ti to V to Mn, an increasing number of up spin $d$ states become occupied, resulting in an increasing magnetic moment. It is interesting to observe that a La-Cr-S compound is known to exist in the same NH$_4$CdCl$_3$ structure. It has been found to show intriguing magnetic properties, which are reminiscent of spin glass behavior.~\cite{lafond1999magnetic} If the compounds that we have predicted in this structure type were synthesized, then one would have a series of compounds crystallizing in the same structure with changing transition metal from Ti to Mn. This can provide a route to study trends in various properties with varying $d$ orbital valence.

\section{Experimental Procedure} 

We carried out solid state and gas flow reactions to attempt bulk synthesis of the predicted compounds. Handling of reagents was performed in a glove box under argon. Traditional \emph{ex situ} reactions were conducted by mixing reagents in a mortar and pestle and then loading them in 16 mm diameter quartz tubes. These tubes were then sealed under vacuum and reacted in box furnaces at high temperatures with a ramp rate of 10$^\circ$C per minute and 8~h hold time.  The reacted tubes were then water quenched at high temperatures to investigate the predicted ternary phases. Lower temperatures were also investigated due to the excess sulfur content in some reactions, and the tendency for ternary phases to form from these liquids at low temperatures.

The vapor pressures of sulfur and selenium put an upper limit to temperatures that can be reached in a solid state reaction, typically around 700$^\circ$C in our work.~\cite{rau1973high,klimova2006investigation,baker1968vapour} CS$_{2}$ has been known to be a powerful sulfidizing agent for oxides, which allows gas flow reactions to be carried out at higher temperatures.~\cite{JACE:JACE1185,saad2010synthesis,hirai1998synthesis} Initial stoichiometric powders were kept in an alumina crucible and then loaded in a quartz reaction tube. The quartz tube was then purged with nitrogen. A mixture of CS$_{2}$ (99.9\%) and nitrogen carrier gas was then flown through the reaction tube while the powders were reacted at high temperatures. 

\textit{Ex situ} powder X-ray diffraction measurements were conducted in reflection with a Bruker D5000 diffractometer with Cu-K$\alpha$ radiation and in transmission with a Bruker D8 diffractometer with Mo-K$\alpha$ radiation. \textit{In situ} high temperature diffraction studies were conducted in transmission with a TC-Transmission capillary stage. The stage consists of an oven which uniformly heats the sample inside a quartz capillary. Here those capillaries are sealed, 0.7 mm diameter, and doubly contained in another capillary of diameter 1 mm for support and containment. The samples were heated at 1$^\circ$C/min, held at their maximum temperature for 8 hours, then cooled at 1$^\circ$C/min. Rietveld analysis was carried out using TOPAS 5.~\cite{coelho2004topas}

Thin film samples of some of the predicted compounds were grown in a multi-source molecular beam epitaxy system optimized for combinatorial composition spread. The substrates were 75 mm diameter and subtended an area large enough to span about 30\% spread in beam flux from each source. For a ternary system, four growths targeting compositions in the middle of the phase diagram and nearer each vertex provides enough range of compositional phase space to adequately survey the compositional phase diagram and determine whether a new phase exists. The growths were carried out at 300$^\circ$C. Selenium was supplied from an effusion source with a valved thermal cracker operating at 1000$^\circ$C. This provided about 75\% of the Se flux in the form of Se$_2$. The films grown were 100 nm thick.

We have attempted synthesis of most of the predicted compounds as summarized in Table~\ref{table_summary}. We did not try to synthesize compounds which have completely empty (Sc based compounds) or completely filled (SrCu$_2$S$_2$ and SrCu$_2$Se$_2$) $d$ orbitals, since we expect that these would not show interesting magnetic properties. Next, we describe our findings for each of the ternary systems that we have investigated.

\subsection{Ca-Mn-S}

\begin{figure*}
	\centering\includegraphics[scale=0.8]{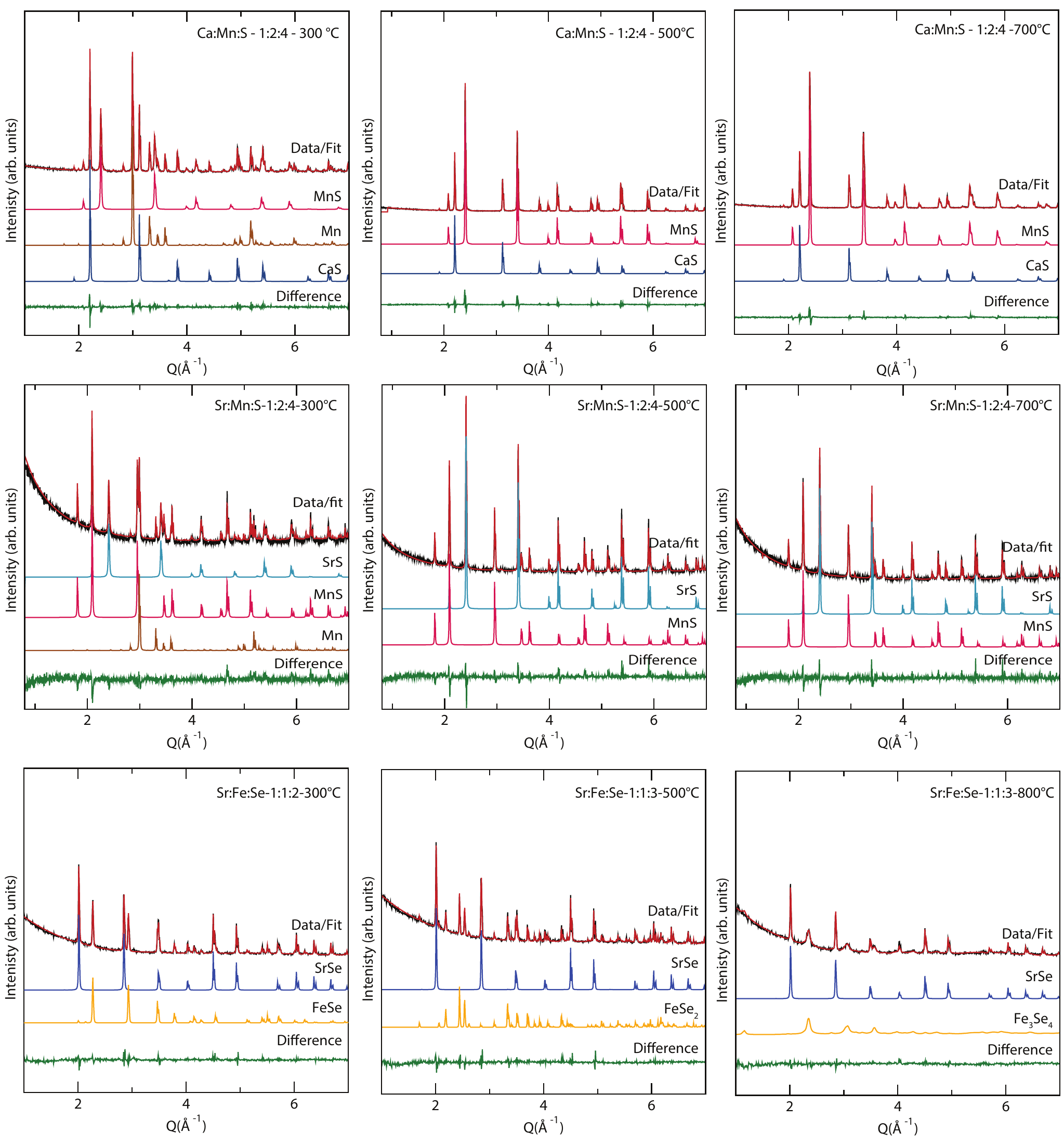} \\
	\caption{(Color online) X-ray powder diffraction patterns of candidate materials for the Ca-Mn-S, Sr-Mn-S and Sr-Fe-Se ternary systems. The reagents were reacted for 16~h and water quenched at temperatures indicated. The data and fit are shown in black and red respectively. Rietveld refinement does not indicate formation of a ternary phase.}
	
	\label{fig:fig21}  
\end{figure*}

CaS (Alfa Aesar, 99.9$\%$), Mn (Alfa Aesar, 99.3$\%$) and S (Alfa Aesar, 99.5$\%$) powders were mixed in stoichiometric ratios and reacted at high temperatures under vacuum. Sulfidation reactions were also carried out using CS$_2$ at 900$^\circ$C. X-ray diffraction and Rietveld analysis shows presence of binary CaS and MnS in the reacted products (Fig.~\ref{fig:fig21}). These results are also summarized in Tables~\ref{exsitu} and~\ref{cs2}. 

An \emph{in situ} study of the initial reaction mixture containing CaS, Mn metal, and S in the ratio of 1:2:3 was also carried out. The powder was loaded in a quartz capillary and sealed under vacuum. 76 diffraction patterns were collected in a duration of 36 hours as the capillary was heated to 700$^\circ$C at the rate of 1$^\circ$C per minute, held at 700$^\circ$C for 8 hours and cooled at the rate of 1$^\circ$C per minute. \emph{In situ} diffraction study further confirms the absence of any ternary phase in the RT-700$^\circ$C temperature range.

\subsection{Sr-Mn-S}

SrS (Strem , 99.9$\%$), Mn and S powders were mixed in stoichiometric ratio in order to check for predicted ternary phases and reacted at various temperatures. X-ray diffraction showed no signs of a ternary phase (Fig.~\ref{fig:fig21}). Sulfidation was also carried using CS$_2$ at 900$^\circ$C. The results of the \emph{ex situ} reactions and gas flow reactions are displayed in Table~\ref{exsitu} and Table~\ref{cs2}  respectively. An \emph{in situ} study was also conducted for 1:2:8 (SrS:Mn:S) composition. The capillary was heated at 1$^\circ$C/min to 500$^\circ$C, held at 500$^\circ$C for 8~h and then cooled at 1$^\circ$C/min. The \emph{in situ} diffraction patterns obtained did not indicate formation of a ternary phase.

  \begin{figure*}
  	\centering\includegraphics[scale=0.8]{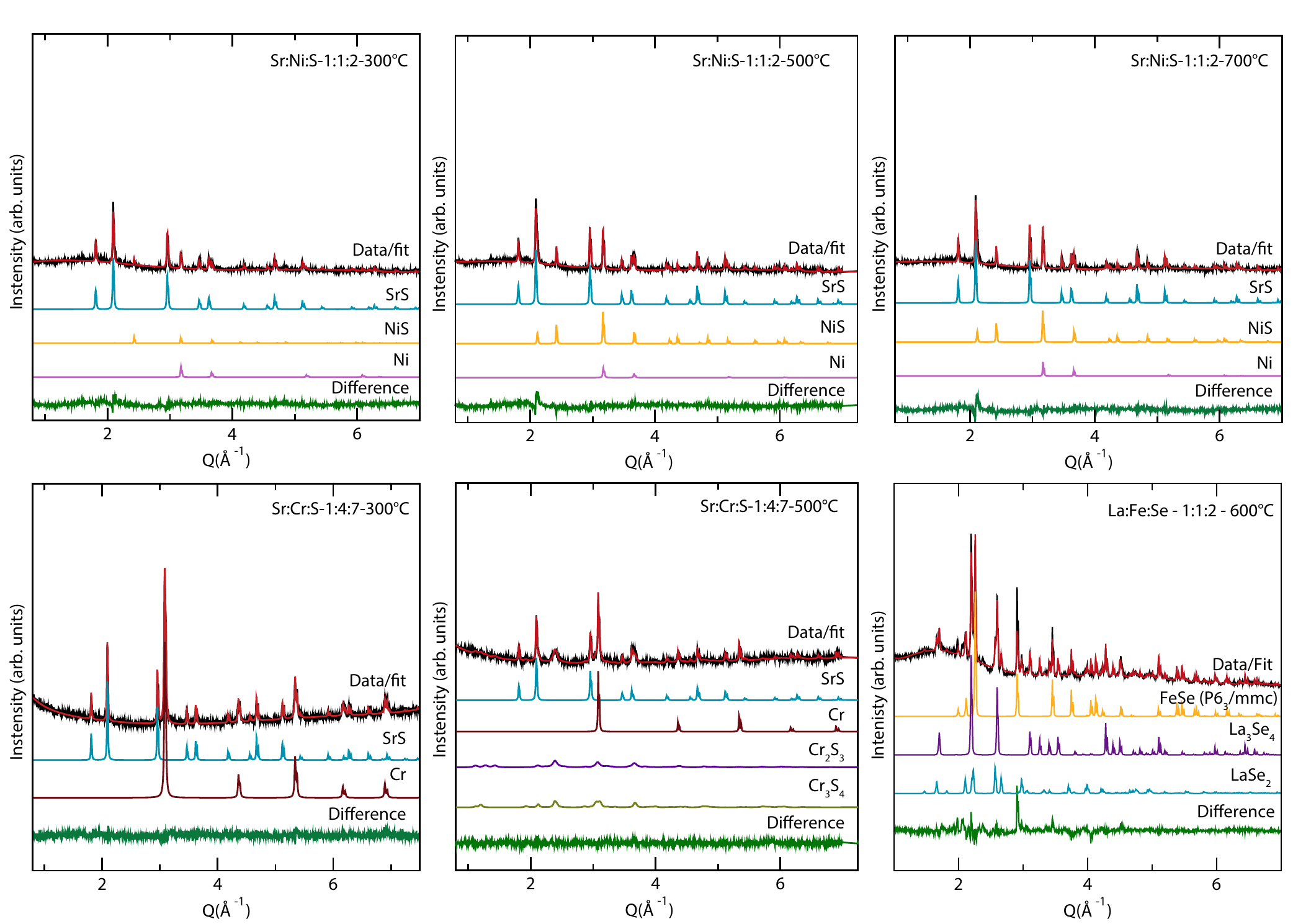} \\
  	\caption{(Color online) X-ray powder diffraction patterns of candidate materials for the Sr-Ni-S, Sr-Cr-S and La-Fe-Se ternary systems. The reagents were reacted for 16~h and water quenched at temperatures indicated. The data and fit are shown in black and red respectively. Rietveld refinement does not indicate formation of a ternary phase.}
  	
  	\label{fig:fig22}  
  \end{figure*}

\subsection{Sr-Fe-Se}

Equimolar Sr chunks (MP Biomedicals, 98$\%$) and Se powder (Alfa Aesar, 99.999$\%$) were reacted using a tube-in-a-tube method\cite{shoemaker2012understanding} to form pure strontium selenide (SrSe). The SrSe precursor was then mixed with Fe and Se powder for carrying out \emph{ex situ} reactions. No ternary compounds were formed (Fig.~\ref{fig:fig21}). The results are displayed in Table~\ref{exsitu}.

\subsection{Sr-Ni-S}
SrS, Ni (Acros, 99$\%$) and S powders were mixed in a stoichiometric ratio of 1:1:1 and reacted at elevated temperatures. No new ternary phase was observed from X-ray diffraction patterns (Fig.~\ref{fig:fig22}).

\subsection{Sr-Cr-S}
SrS, Cr (Alfa Aesar, 99$\%$) and S powders were mixed in 1:4:6 stoichiometric ratio and reacted at high temperatures. X-ray diffraction did not indicate formation of a ternary phase (Fig.~\ref{fig:fig22}).

\subsection{La-Fe-Se}    

La (Alfa Aesar 99.9 $\%$ pure), Fe and Se powder were used to carry out \emph{ex situ} reactions, which resulted in the formation of binaries (Fig.~\ref{fig:fig22}). The results are displayed in Table~\ref{exsitu}.

\subsection{Sr-V-S}

SrS, V (Strem, 99$\%$), and S powders were mixed in stoichiometric ratios and reacted at high temperatures. The results are shown in Table~\ref{exsitu}. The measured X-ray powder diffraction patterns of the quenched samples were complex due to the presence of multiple incompletely-reacted binary vanadium sulfides and SrS. A satisfactory Rietveld fit could not be obtained, but, the diffraction patterns did not contain any unindexed peaks. No low-angle peaks from the predicted Sr$_2$VS$_4$ or SrV$_4$S$_7$ phases were seen. An \emph{in situ} study was also conducted for 1:4:7 (SrS:Cr:S) composition. The capillary was heated at 1$^\circ$C/min to 500$^\circ$C, held at 500$^\circ$C for 8~h and then cooled at 1$^\circ$C/min. The \emph{in situ} diffraction patterns obtained did not indicate formation of a ternary phase.

\subsection{Pb-Fe-Se}

\begin{figure}
	\centering\includegraphics[scale=0.50]{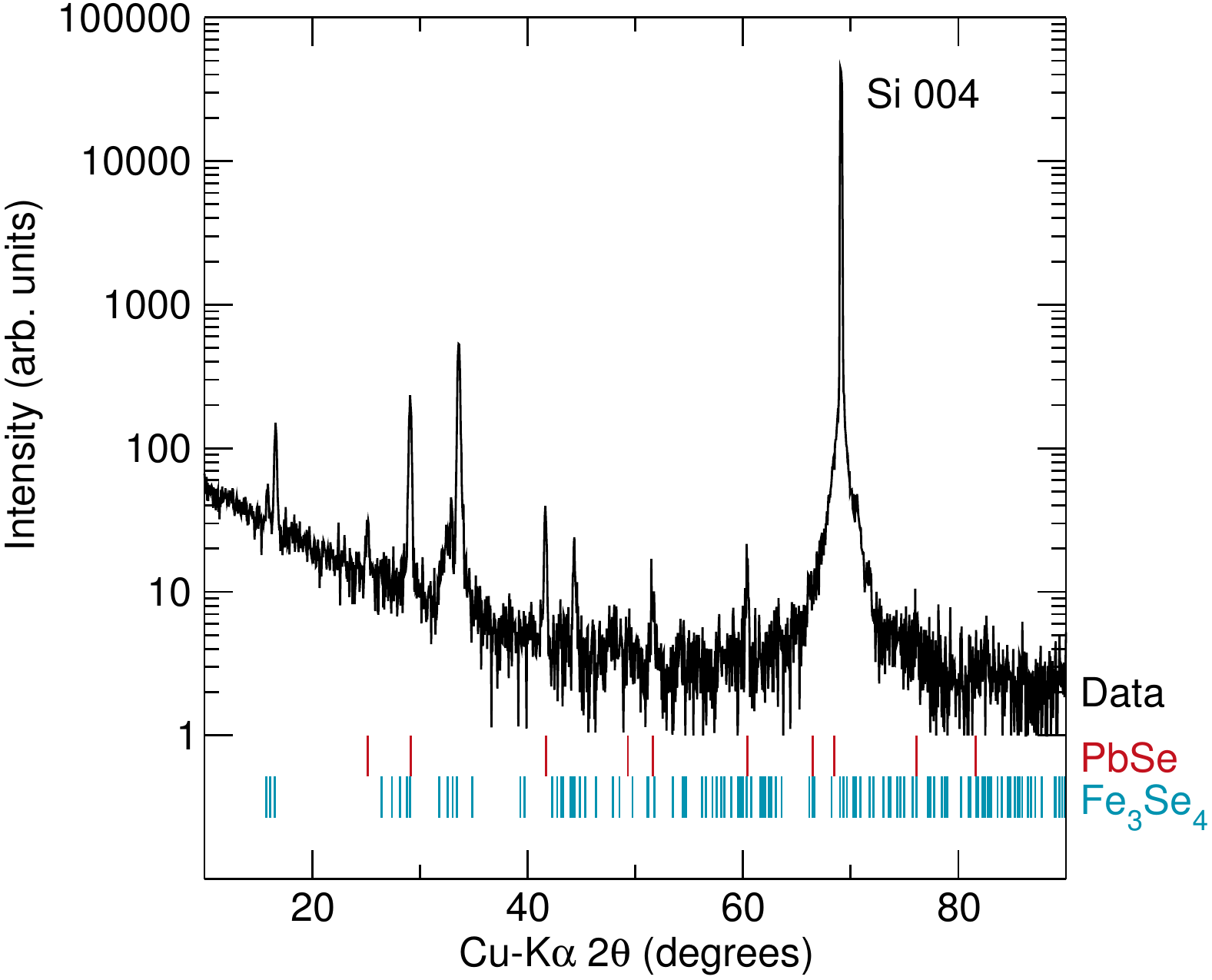} \\
	\caption{Bragg peaks in the X-ray diffraction pattern of a Pb-Fe-Se film grown by MBE,
	can be indexed using a combination of PbSe, Fe$_3$Se$_4$,
	and the Si substrate. No new peaks that would correspond to the predicted phases in Table \ref{table_summary} are seen.
	}
	\label{fig:PbFeSe}  
\end{figure}

Thin film deposition can often stabilize phases that are not seen in the bulk due to interfacial effects and control of chemical potential of some sources as well as the fact that synthesis can be carried out at much lower temperatures. Thin films of Pb-Fe-Se were grown searching for the predicted composition PbFe$_2$Se$_3$. They exhibited the diffraction pattern shown in Figure \ref{fig:PbFeSe}. The oriented nature of the sample and the small thickness preclude a meaningful Rietveld fit of the data, but all peaks are accounted for by a combination of PbSe, Fe$_3$Se$_4$, and the Si substrate. No peaks from our predicted ternary Pb-Fe-Se phases PbFe$_2$Se$_3$, Pb$_2$FeSe$_3$, or Pb$_2$FeSe$_4$ nor from FeSe were seen. It is interesting to note that the iron selenide phase that was formed has a larger Se content than FeSe. We have found that this is determined by the Se$_2$ flux from the cracked Se source. Unlike bulk synthesis where the number of Se atoms is determined, in thin film growth the beam composition and flux determines the Se chemical potential. For large fluxes of more reactive molecules the chemical potential is larger and the composition is pushed to large Se content.

\begin{table*}

	\centering
	\caption{Summary of direct combination \emph{ex situ} reactions}
	\begin{tabular}{|>{\centering\arraybackslash} m{2cm} |>{\centering\arraybackslash} m{2.25cm} | >{\centering\arraybackslash}m{1.75cm}|>{\centering\arraybackslash}m{4cm}|>{\centering\arraybackslash}m{4.6cm}|>{\centering\arraybackslash}m{2.7cm}|}
		\hline
		\multicolumn{6}{|c|}{\textbf{Sr-Mn-S}} \\ \hline
		
		Reagents & Molar ratio SrS:Mn:S & Sr:Mn:S & 300$^\circ$C products & 500$^\circ$C products& 700$^\circ$C products\\ \hline
		SrS, Mn, S & 2:1:4 & 2:1:6 & SrS + MnS + Mn + S & SrS + MnS + S &   \\ 
		SrS, Mn, S& 1:2:5 & 1:2:6 & SrS + MnS + Mn + S & SrS + MnS + S & \\ 
		SrS, Mn, S& 2:1:6 & 2:1:8 & SrS + MnS + Mn + S & SrS + MnS + S &\\ 
		SrS, Mn, S& 1:2:7 & 1:2:8 & SrS + MnS + Mn + S & SrS + MnS + S &  \\ 
		SrS, Mn, S& 1:2:3 & 1:2:4& SrS + MnS + Mn + S & SrS + MnS & SrS+MnS \\ \hline
		\multicolumn{6}{|c|}{\textbf{Ca-Mn-S}} \\ \hline
		Reagents& Molar ratio CaS:Mn:S&  Ca:Mn:S  & 300$^\circ$C products & 500$^\circ$C products & 700$^\circ$C products\\ \hline
		CaS, Mn, S  & 1:2:3 & 1:2:4 & CaS + MnS + Mn & CaS + MnS & CaS + MnS \\ \hline
		\multicolumn{6}{|c|}{\textbf{Sr-Ni-S}} \\ \hline
		Reagents& Molar ratio SrS:Ni:S & Sr:Ni:S & 300$^\circ$C products & 500$^\circ$C products & 700$^\circ$C products\\ \hline
		SrS, Ni, S  & 1:1:1 & 1:1:2 & SrS + Ni + NiS & SrS + Ni + NiS & SrS + Ni + NiS  \\ \hline
		\multicolumn{6}{|c|}{\textbf{Sr-Cr-S}} \\ \hline
		Reagents& Molar ratio SrS:Cr:S & Sr:Cr:S & 300$^\circ$C products & 500$^\circ$C products & \\ \hline
		SrS, Ni, S  & 1:4:6 & 1:4:7 & SrS + Cr & SrS + Cr + Cr$_2$S$_3$ + Cr$_3$S$_4$ & \\ \hline
		\multicolumn{6}{|c|}{\textbf{Sr-Fe-Se}} \\ \hline
		Reagents& Molar ratio SrSe:Fe:Se & Sr:Fe:Se & & 500$^\circ$C products & 800$^\circ$C products  \\ \hline
		SrSe, Fe, Se & 1:1:1 & 1:1:2 & & SrSe + FeSe & \emph{in situ} \\ 
		SrSe, Fe, Se& 1:1:2 & 1:1:3 & & SrSe + FeSe$_2$ & SrSe + Fe$_3$Se$_4$ \\ \hline

		\multicolumn{6}{|c|}{\textbf{La-Fe-Se}} \\ \hline
		Reagents& Molar ratio La:Fe:Se & La:Fe:Se & & 600$^\circ$C products  & \\ \hline
		La, Fe, Se & 1:1:2 & 1:1:2 &  & FeSe + La$_{3}$Se$_4$ + LaSe$_2$ &   \\ \hline

	\end{tabular} \label{exsitu}
	

\end{table*}

%

		
		

	

\begin{table*}[]

	\centering
	\caption{Summary of CS$_2$ gas flow reactions}
	\begin{tabular}{|>{\centering\arraybackslash} m{5cm} |>{\centering\arraybackslash} m{4cm} | >{\centering\arraybackslash}m{5cm}|}
		\hline
		\multicolumn{3}{|c|}{\textbf{Sr-Mn-S}} \\
		\hline
		Reagents & Molar ratio Sr:Mn & 900$^\circ$C \\ \hline
		SrCO$_3$, Mn & 1:2 & SrS + MnS  \\ \hline
		
		\multicolumn{3}{|c|}{\textbf{Ca-Mn-S}} \\
		\hline
		Reagents & Molar ratio Ca:Mn & 900$^\circ$C  \\ \hline
		CaCO$_3$, Mn & 1:2 & CaS + MnS    \\ \hline
		
		
		\multicolumn{3}{|c|}{\textbf{Sr-V-S}} \\
		\hline
		Reagents & Molar ratio Sr:V & 900$^\circ$C  \\ \hline
		SrCO$_3$, V & 3:5 & SrS + V$_3$S$_4$     \\ \hline
		
		
		\multicolumn{3}{|c|}{\textbf{Sr-Ni-S}} \\
		\hline
		Reagents & Molar ratio Sr:Ni & 900$^\circ$C  \\ \hline
		SrCO$_3$, Ni & 1:1 & SrS + NiS + NiS$_2$ + Ni$_3$S$_2$    \\ \hline
		
	\end{tabular}	\label{cs2}
	

\end{table*}

\section{Summary and outlook}\label{summary}

To summarize, in this study we investigated transition metal selenide and sulfide ternary systems of the form XYZ, where the cation X= Ba, Ca, Sr, La, K, Bi, Pb, Y is a $3d$ transition metal and Z= S or Se, whose ternary phase diagrams are known to be empty. 
Using a combination of data-mining based ionic substitution model and first-principles density functional theory calculations, we explored several thousand candidate structures and examined them for their stability.  
Using a Bayesian likelihood model to minimize false negative predictions, we narrowed down the search to 24 candidates.
We then carried out high-throughput experiments to attempt synthesis of these compounds using solid state and gas flow reactions. 
While we found similar compounds to known misfit materials, the truly new materials were not found in our experimental synthesis.
There are three possible conclusions that result from this study:
\begin{enumerate}
\item 24 of the ternary phase diagrams are actually empty.
\item There are new structures or structures that do not follow ionic substitution rules in the 24 potentially empty phase diagrams.
\item The synthesis techniques we used were not appropriate for the predicted materials.	
\end{enumerate}
Possibilities 2 and 3 are not mutually exclusive; there may be some new structures and some difficult-to-synthesize compounds in the chemical phase space considered.

Our work can open avenues for further investigations: experimental synthesis attempts of our
proposed candidate materials using alternative synthesis routes will definitely be a direction worth pursuing. Given the interesting correlated electron properties of known chalcogenides, including a recently proposed oxysulfide superconductor~\cite{0295-5075-111-1-17002}, this can be fruitful. Along the theoretical front, future work can include using evolutionary algorithms to find other candidate structures in this interesting chemical space and using more accurate electronic structure methods to check the stability of our proposed compounds.

The work presented here represents a tight coupling of high-throughput quantum calculations with high-throughput experiment and contains several important lessons.
We have shown that inherent errors in the density functional are sufficient to explain failures to synthesize structures that are predicted to be stable using density functional theory calculations.
Higher accuracy calculations such as quantum Monte Carlo (QMC)\cite{wagner2007transition} may be worth the additional computational cost, since they should allow reduction of both the false positive and false negative rates by significant margins.
However, given that currently our high-throughput synthesis techniques are faster than QMC calculations, it may be more optimal to simply accept a larger false positive rate and utilize experiments to winnow candidates.

\section*{Acknowledgments}

We acknowledge support from the Center for Emergent Superconductivity, an Energy Frontier Research Center funded by the U.S. Department of Energy, Office of Science, Office of Basic Energy Sciences under Award Number DEAC0298CH1088. Illuminating discussions with David Ceperley, Laura Greene, Mercouri Kanatzidis, Cedomir Petrovic, John Tranquada and Jian-Min Zuo are gratefully acknowledged. Computational resources were provided by the University of Illinois Campus Cluster. \emph{Ex situ} initial X-ray diffraction measurements were carried out in part in the Frederick Seitz Materials Research Laboratory Central Research Facilities, University of Illinois. 


\begin{thebibliography}{36}%
\makeatletter
\providecommand \@ifxundefined [1]{%
 \@ifx{#1\undefined}
}%
\providecommand \@ifnum [1]{%
 \ifnum #1\expandafter \@firstoftwo
 \else \expandafter \@secondoftwo
 \fi
}%
\providecommand \@ifx [1]{%
 \ifx #1\expandafter \@firstoftwo
 \else \expandafter \@secondoftwo
 \fi
}%
\providecommand \natexlab [1]{#1}%
\providecommand \enquote  [1]{``#1''}%
\providecommand \bibnamefont  [1]{#1}%
\providecommand \bibfnamefont [1]{#1}%
\providecommand \citenamefont [1]{#1}%
\providecommand \href@noop [0]{\@secondoftwo}%
\providecommand \href [0]{\begingroup \@sanitize@url \@href}%
\providecommand \@href[1]{\@@startlink{#1}\@@href}%
\providecommand \@@href[1]{\endgroup#1\@@endlink}%
\providecommand \@sanitize@url [0]{\catcode `\\12\catcode `\$12\catcode
  `\&12\catcode `\#12\catcode `\^12\catcode `\_12\catcode `\%12\relax}%
\providecommand \@@startlink[1]{}%
\providecommand \@@endlink[0]{}%
\providecommand \url  [0]{\begingroup\@sanitize@url \@url }%
\providecommand \@url [1]{\endgroup\@href {#1}{\urlprefix }}%
\providecommand \urlprefix  [0]{URL }%
\providecommand \Eprint [0]{\href }%
\providecommand \doibase [0]{http://dx.doi.org/}%
\providecommand \selectlanguage [0]{\@gobble}%
\providecommand \bibinfo  [0]{\@secondoftwo}%
\providecommand \bibfield  [0]{\@secondoftwo}%
\providecommand \translation [1]{[#1]}%
\providecommand \BibitemOpen [0]{}%
\providecommand \bibitemStop [0]{}%
\providecommand \bibitemNoStop [0]{.\EOS\space}%
\providecommand \EOS [0]{\spacefactor3000\relax}%
\providecommand \BibitemShut  [1]{\csname bibitem#1\endcsname}%
\let\auto@bib@innerbib\@empty
\bibitem [{\citenamefont {Curtarolo}\ \emph {et~al.}(2013)\citenamefont
  {Curtarolo}, \citenamefont {Hart}, \citenamefont {Nardelli}, \citenamefont
  {Mingo}, \citenamefont {Sanvito},\ and\ \citenamefont
  {Levy}}]{curtarolo2013high}%
  \BibitemOpen
  \bibfield  {author} {\bibinfo {author} {\bibfnamefont {Stefano}\ \bibnamefont
  {Curtarolo}}, \bibinfo {author} {\bibfnamefont {Gus~LW}\ \bibnamefont
  {Hart}}, \bibinfo {author} {\bibfnamefont {Marco~Buongiorno}\ \bibnamefont
  {Nardelli}}, \bibinfo {author} {\bibfnamefont {Natalio}\ \bibnamefont
  {Mingo}}, \bibinfo {author} {\bibfnamefont {Stefano}\ \bibnamefont
  {Sanvito}}, \ and\ \bibinfo {author} {\bibfnamefont {Ohad}\ \bibnamefont
  {Levy}},\ }\bibfield  {title} {\enquote {\bibinfo {title} {The
  high-throughput highway to computational materials design},}\ }\href@noop {}
  {\bibfield  {journal} {\bibinfo  {journal} {Nature materials}\ }\textbf
  {\bibinfo {volume} {12}},\ \bibinfo {pages} {191--201} (\bibinfo {year}
  {2013})}\BibitemShut {NoStop}%
\bibitem [{\citenamefont {Jain}\ \emph {et~al.}(2013)\citenamefont {Jain},
  \citenamefont {Ong}, \citenamefont {Hautier}, \citenamefont {Chen},
  \citenamefont {Richards}, \citenamefont {Dacek}, \citenamefont {Cholia},
  \citenamefont {Gunter}, \citenamefont {Skinner}, \citenamefont {Ceder} \emph
  {et~al.}}]{jain2013commentary}%
  \BibitemOpen
  \bibfield  {author} {\bibinfo {author} {\bibfnamefont {Anubhav}\ \bibnamefont
  {Jain}}, \bibinfo {author} {\bibfnamefont {Shyue~Ping}\ \bibnamefont {Ong}},
  \bibinfo {author} {\bibfnamefont {Geoffroy}\ \bibnamefont {Hautier}},
  \bibinfo {author} {\bibfnamefont {Wei}\ \bibnamefont {Chen}}, \bibinfo
  {author} {\bibfnamefont {William~Davidson}\ \bibnamefont {Richards}},
  \bibinfo {author} {\bibfnamefont {Stephen}\ \bibnamefont {Dacek}}, \bibinfo
  {author} {\bibfnamefont {Shreyas}\ \bibnamefont {Cholia}}, \bibinfo {author}
  {\bibfnamefont {Dan}\ \bibnamefont {Gunter}}, \bibinfo {author}
  {\bibfnamefont {David}\ \bibnamefont {Skinner}}, \bibinfo {author}
  {\bibfnamefont {Gerbrand}\ \bibnamefont {Ceder}},  \emph {et~al.},\
  }\bibfield  {title} {\enquote {\bibinfo {title} {Commentary: The materials
  project: A materials genome approach to accelerating materials innovation},}\
  }\href@noop {} {\bibfield  {journal} {\bibinfo  {journal} {APL Materials}\
  }\textbf {\bibinfo {volume} {1}},\ \bibinfo {pages} {011002} (\bibinfo {year}
  {2013})}\BibitemShut {NoStop}%
\bibitem [{\citenamefont {Curtarolo}\ \emph {et~al.}(2012)\citenamefont
  {Curtarolo}, \citenamefont {Setyawan}, \citenamefont {Wang}, \citenamefont
  {Xue}, \citenamefont {Yang}, \citenamefont {Taylor}, \citenamefont {Nelson},
  \citenamefont {Hart}, \citenamefont {Sanvito}, \citenamefont
  {Buongiorno-Nardelli} \emph {et~al.}}]{curtarolo2012aflowlib}%
  \BibitemOpen
  \bibfield  {author} {\bibinfo {author} {\bibfnamefont {Stefano}\ \bibnamefont
  {Curtarolo}}, \bibinfo {author} {\bibfnamefont {Wahyu}\ \bibnamefont
  {Setyawan}}, \bibinfo {author} {\bibfnamefont {Shidong}\ \bibnamefont
  {Wang}}, \bibinfo {author} {\bibfnamefont {Junkai}\ \bibnamefont {Xue}},
  \bibinfo {author} {\bibfnamefont {Kesong}\ \bibnamefont {Yang}}, \bibinfo
  {author} {\bibfnamefont {Richard~H}\ \bibnamefont {Taylor}}, \bibinfo
  {author} {\bibfnamefont {Lance~J}\ \bibnamefont {Nelson}}, \bibinfo {author}
  {\bibfnamefont {Gus~LW}\ \bibnamefont {Hart}}, \bibinfo {author}
  {\bibfnamefont {Stefano}\ \bibnamefont {Sanvito}}, \bibinfo {author}
  {\bibfnamefont {Marco}\ \bibnamefont {Buongiorno-Nardelli}},  \emph
  {et~al.},\ }\bibfield  {title} {\enquote {\bibinfo {title} {Aflowlib. org: A
  distributed materials properties repository from high-throughput ab initio
  calculations},}\ }\href@noop {} {\bibfield  {journal} {\bibinfo  {journal}
  {Computational Materials Science}\ }\textbf {\bibinfo {volume} {58}},\
  \bibinfo {pages} {227--235} (\bibinfo {year} {2012})}\BibitemShut {NoStop}%
\bibitem [{\citenamefont {Saal}\ \emph {et~al.}(2013)\citenamefont {Saal},
  \citenamefont {Kirklin}, \citenamefont {Aykol}, \citenamefont {Meredig},\
  and\ \citenamefont {Wolverton}}]{saal2013materials}%
  \BibitemOpen
  \bibfield  {author} {\bibinfo {author} {\bibfnamefont {James~E}\ \bibnamefont
  {Saal}}, \bibinfo {author} {\bibfnamefont {Scott}\ \bibnamefont {Kirklin}},
  \bibinfo {author} {\bibfnamefont {Muratahan}\ \bibnamefont {Aykol}}, \bibinfo
  {author} {\bibfnamefont {Bryce}\ \bibnamefont {Meredig}}, \ and\ \bibinfo
  {author} {\bibfnamefont {Christopher}\ \bibnamefont {Wolverton}},\ }\bibfield
   {title} {\enquote {\bibinfo {title} {Materials design and discovery with
  high-throughput density functional theory: The open quantum materials
  database (oqmd)},}\ }\href@noop {} {\bibfield  {journal} {\bibinfo  {journal}
  {Jom}\ }\textbf {\bibinfo {volume} {65}},\ \bibinfo {pages} {1501--1509}
  (\bibinfo {year} {2013})}\BibitemShut {NoStop}%
\bibitem [{com()}]{compmatrep}%
  \BibitemOpen
  \href@noop {} {\enquote {\bibinfo {title} {Computational materials
  repository},}\ }\bibinfo {howpublished}
  {\url{https://cmr.fysik.dtu.dk/}}\BibitemShut {NoStop}%
\bibitem [{ele()}]{elecstrucproj}%
  \BibitemOpen
  \href@noop {} {\enquote {\bibinfo {title} {Electronic structure project},}\
  }\bibinfo {howpublished} {\url{http://gurka.fysik.uu.se/ESP/}}\BibitemShut
  {NoStop}%
\bibitem [{\citenamefont {Hautier}\ \emph
  {et~al.}(2010{\natexlab{a}})\citenamefont {Hautier}, \citenamefont {Fischer},
  \citenamefont {Jain}, \citenamefont {Mueller},\ and\ \citenamefont
  {Ceder}}]{hautier2010finding}%
  \BibitemOpen
  \bibfield  {author} {\bibinfo {author} {\bibfnamefont {Geoffroy}\
  \bibnamefont {Hautier}}, \bibinfo {author} {\bibfnamefont {Christopher~C}\
  \bibnamefont {Fischer}}, \bibinfo {author} {\bibfnamefont {Anubhav}\
  \bibnamefont {Jain}}, \bibinfo {author} {\bibfnamefont {Tim}\ \bibnamefont
  {Mueller}}, \ and\ \bibinfo {author} {\bibfnamefont {Gerbrand}\ \bibnamefont
  {Ceder}},\ }\bibfield  {title} {\enquote {\bibinfo {title} {Finding
  nature’s missing ternary oxide compounds using machine learning and density
  functional theory},}\ }\href@noop {} {\bibfield  {journal} {\bibinfo
  {journal} {Chemistry of Materials}\ }\textbf {\bibinfo {volume} {22}},\
  \bibinfo {pages} {3762--3767} (\bibinfo {year}
  {2010}{\natexlab{a}})}\BibitemShut {NoStop}%
\bibitem [{\citenamefont {Hautier}\ \emph {et~al.}(2011)\citenamefont
  {Hautier}, \citenamefont {Jain}, \citenamefont {Ong}, \citenamefont {Kang},
  \citenamefont {Moore}, \citenamefont {Doe},\ and\ \citenamefont
  {Ceder}}]{hautier2011phosphates}%
  \BibitemOpen
  \bibfield  {author} {\bibinfo {author} {\bibfnamefont {Geoffroy}\
  \bibnamefont {Hautier}}, \bibinfo {author} {\bibfnamefont {Anubhav}\
  \bibnamefont {Jain}}, \bibinfo {author} {\bibfnamefont {Shyue~Ping}\
  \bibnamefont {Ong}}, \bibinfo {author} {\bibfnamefont {Byoungwoo}\
  \bibnamefont {Kang}}, \bibinfo {author} {\bibfnamefont {Charles}\
  \bibnamefont {Moore}}, \bibinfo {author} {\bibfnamefont {Robert}\
  \bibnamefont {Doe}}, \ and\ \bibinfo {author} {\bibfnamefont {Gerbrand}\
  \bibnamefont {Ceder}},\ }\bibfield  {title} {\enquote {\bibinfo {title}
  {Phosphates as lithium-ion battery cathodes: an evaluation based on
  high-throughput ab initio calculations},}\ }\href@noop {} {\bibfield
  {journal} {\bibinfo  {journal} {Chemistry of Materials}\ }\textbf {\bibinfo
  {volume} {23}},\ \bibinfo {pages} {3495--3508} (\bibinfo {year}
  {2011})}\BibitemShut {NoStop}%
\bibitem [{\citenamefont {Chen}\ \emph {et~al.}(2012)\citenamefont {Chen},
  \citenamefont {Hautier},\ and\ \citenamefont {Ceder}}]{chen2012synthesis}%
  \BibitemOpen
  \bibfield  {author} {\bibinfo {author} {\bibfnamefont {Hailong}\ \bibnamefont
  {Chen}}, \bibinfo {author} {\bibfnamefont {Geoffroy}\ \bibnamefont
  {Hautier}}, \ and\ \bibinfo {author} {\bibfnamefont {Gerbrand}\ \bibnamefont
  {Ceder}},\ }\bibfield  {title} {\enquote {\bibinfo {title} {Synthesis,
  computed stability, and crystal structure of a new family of inorganic
  compounds: carbonophosphates},}\ }\href@noop {} {\bibfield  {journal}
  {\bibinfo  {journal} {Journal of the American Chemical Society}\ }\textbf
  {\bibinfo {volume} {134}},\ \bibinfo {pages} {19619--19627} (\bibinfo {year}
  {2012})}\BibitemShut {NoStop}%
\bibitem [{\citenamefont {Hautier}\ \emph {et~al.}(2013)\citenamefont
  {Hautier}, \citenamefont {Miglio}, \citenamefont {Ceder}, \citenamefont
  {Rignanese},\ and\ \citenamefont {Gonze}}]{hautier2013identification}%
  \BibitemOpen
  \bibfield  {author} {\bibinfo {author} {\bibfnamefont {Geoffroy}\
  \bibnamefont {Hautier}}, \bibinfo {author} {\bibfnamefont {Anna}\
  \bibnamefont {Miglio}}, \bibinfo {author} {\bibfnamefont {Gerbrand}\
  \bibnamefont {Ceder}}, \bibinfo {author} {\bibfnamefont {Gian-Marco}\
  \bibnamefont {Rignanese}}, \ and\ \bibinfo {author} {\bibfnamefont {Xavier}\
  \bibnamefont {Gonze}},\ }\bibfield  {title} {\enquote {\bibinfo {title}
  {Identification and design principles of low hole effective mass p-type
  transparent conducting oxides},}\ }\href@noop {} {\bibfield  {journal}
  {\bibinfo  {journal} {Nature communications}\ }\textbf {\bibinfo {volume}
  {4}} (\bibinfo {year} {2013})}\BibitemShut {NoStop}%
\bibitem [{\citenamefont {Cerqueira}\ \emph {et~al.}(2015)\citenamefont
  {Cerqueira}, \citenamefont {Lin}, \citenamefont {Amsler}, \citenamefont
  {Goedecker}, \citenamefont {Botti},\ and\ \citenamefont
  {Marques}}]{cerqueira2015identification}%
  \BibitemOpen
  \bibfield  {author} {\bibinfo {author} {\bibfnamefont {Tiago~FT}\
  \bibnamefont {Cerqueira}}, \bibinfo {author} {\bibfnamefont {Sun}\
  \bibnamefont {Lin}}, \bibinfo {author} {\bibfnamefont {Maximilian}\
  \bibnamefont {Amsler}}, \bibinfo {author} {\bibfnamefont {Stefan}\
  \bibnamefont {Goedecker}}, \bibinfo {author} {\bibfnamefont {Silvana}\
  \bibnamefont {Botti}}, \ and\ \bibinfo {author} {\bibfnamefont {Miguel~AL}\
  \bibnamefont {Marques}},\ }\bibfield  {title} {\enquote {\bibinfo {title}
  {Identification of novel cu, ag, and au ternary oxides from global structural
  prediction},}\ }\href@noop {} {\bibfield  {journal} {\bibinfo  {journal}
  {arXiv preprint arXiv:1503.07327}\ } (\bibinfo {year} {2015})}\BibitemShut
  {NoStop}%
\bibitem [{\citenamefont {Hautier}\ \emph
  {et~al.}(2010{\natexlab{b}})\citenamefont {Hautier}, \citenamefont {Fischer},
  \citenamefont {Ehrlacher}, \citenamefont {Jain},\ and\ \citenamefont
  {Ceder}}]{hautier2010data}%
  \BibitemOpen
  \bibfield  {author} {\bibinfo {author} {\bibfnamefont {Geoffroy}\
  \bibnamefont {Hautier}}, \bibinfo {author} {\bibfnamefont {Chris}\
  \bibnamefont {Fischer}}, \bibinfo {author} {\bibfnamefont {Virginie}\
  \bibnamefont {Ehrlacher}}, \bibinfo {author} {\bibfnamefont {Anubhav}\
  \bibnamefont {Jain}}, \ and\ \bibinfo {author} {\bibfnamefont {Gerbrand}\
  \bibnamefont {Ceder}},\ }\bibfield  {title} {\enquote {\bibinfo {title} {Data
  mined ionic substitutions for the discovery of new compounds},}\ }\href@noop
  {} {\bibfield  {journal} {\bibinfo  {journal} {Inorganic chemistry}\ }\textbf
  {\bibinfo {volume} {50}},\ \bibinfo {pages} {656--663} (\bibinfo {year}
  {2010}{\natexlab{b}})}\BibitemShut {NoStop}%
\bibitem [{ics()}]{icsd}%
  \BibitemOpen
  \href@noop {} {\enquote {\bibinfo {title} {Inorganic crystal structure
  database},}\ }\bibinfo {howpublished}
  {\url{http://www.fiz-karlsruhe.de/icsd.html}}\BibitemShut {NoStop}%
\bibitem [{\citenamefont {Hundt}\ \emph {et~al.}(2006)\citenamefont {Hundt},
  \citenamefont {Sch{\"o}n},\ and\ \citenamefont {Jansen}}]{hundt2006cmpz}%
  \BibitemOpen
  \bibfield  {author} {\bibinfo {author} {\bibfnamefont {R}~\bibnamefont
  {Hundt}}, \bibinfo {author} {\bibfnamefont {JC}~\bibnamefont {Sch{\"o}n}}, \
  and\ \bibinfo {author} {\bibfnamefont {M}~\bibnamefont {Jansen}},\ }\bibfield
   {title} {\enquote {\bibinfo {title} {Cmpz--an algorithm for the efficient
  comparison of periodic structures},}\ }\href@noop {} {\bibfield  {journal}
  {\bibinfo  {journal} {Journal of applied crystallography}\ }\textbf {\bibinfo
  {volume} {39}},\ \bibinfo {pages} {6--16} (\bibinfo {year}
  {2006})}\BibitemShut {NoStop}%
\bibitem [{\citenamefont {Oganov}\ and\ \citenamefont
  {Glass}(2006)}]{oganov2006crystal}%
  \BibitemOpen
  \bibfield  {author} {\bibinfo {author} {\bibfnamefont {Artem~R}\ \bibnamefont
  {Oganov}}\ and\ \bibinfo {author} {\bibfnamefont {Colin~W}\ \bibnamefont
  {Glass}},\ }\bibfield  {title} {\enquote {\bibinfo {title} {Crystal structure
  prediction using ab initio evolutionary techniques: Principles and
  applications},}\ }\href@noop {} {\bibfield  {journal} {\bibinfo  {journal}
  {The Journal of chemical physics}\ }\textbf {\bibinfo {volume} {124}},\
  \bibinfo {pages} {244704} (\bibinfo {year} {2006})}\BibitemShut {NoStop}%
\bibitem [{\citenamefont {Trimarchi}\ and\ \citenamefont
  {Zunger}(2007)}]{trimarchi2007global}%
  \BibitemOpen
  \bibfield  {author} {\bibinfo {author} {\bibfnamefont {Giancarlo}\
  \bibnamefont {Trimarchi}}\ and\ \bibinfo {author} {\bibfnamefont {Alex}\
  \bibnamefont {Zunger}},\ }\bibfield  {title} {\enquote {\bibinfo {title}
  {Global space-group optimization problem: Finding the stablest crystal
  structure without constraints},}\ }\href@noop {} {\bibfield  {journal}
  {\bibinfo  {journal} {Physical Review B}\ }\textbf {\bibinfo {volume} {75}},\
  \bibinfo {pages} {104113} (\bibinfo {year} {2007})}\BibitemShut {NoStop}%
\bibitem [{\citenamefont {Giannozzi}\ \emph {et~al.}(2009)\citenamefont
  {Giannozzi}, \citenamefont {Baroni}, \citenamefont {Bonini}, \citenamefont
  {Calandra}, \citenamefont {Car}, \citenamefont {Cavazzoni}, \citenamefont
  {Ceresoli}, \citenamefont {Chiarotti}, \citenamefont {Cococcioni},
  \citenamefont {Dabo} \emph {et~al.}}]{giannozzi2009quantum}%
  \BibitemOpen
  \bibfield  {author} {\bibinfo {author} {\bibfnamefont {Paolo}\ \bibnamefont
  {Giannozzi}}, \bibinfo {author} {\bibfnamefont {Stefano}\ \bibnamefont
  {Baroni}}, \bibinfo {author} {\bibfnamefont {Nicola}\ \bibnamefont {Bonini}},
  \bibinfo {author} {\bibfnamefont {Matteo}\ \bibnamefont {Calandra}}, \bibinfo
  {author} {\bibfnamefont {Roberto}\ \bibnamefont {Car}}, \bibinfo {author}
  {\bibfnamefont {Carlo}\ \bibnamefont {Cavazzoni}}, \bibinfo {author}
  {\bibfnamefont {Davide}\ \bibnamefont {Ceresoli}}, \bibinfo {author}
  {\bibfnamefont {Guido~L}\ \bibnamefont {Chiarotti}}, \bibinfo {author}
  {\bibfnamefont {Matteo}\ \bibnamefont {Cococcioni}}, \bibinfo {author}
  {\bibfnamefont {Ismaila}\ \bibnamefont {Dabo}},  \emph {et~al.},\ }\bibfield
  {title} {\enquote {\bibinfo {title} {Quantum espresso: a modular and
  open-source software project for quantum simulations of materials},}\
  }\href@noop {} {\bibfield  {journal} {\bibinfo  {journal} {Journal of
  Physics: Condensed Matter}\ }\textbf {\bibinfo {volume} {21}},\ \bibinfo
  {pages} {395502} (\bibinfo {year} {2009})}\BibitemShut {NoStop}%
\bibitem [{\citenamefont {Perdew}\ \emph {et~al.}(1996)\citenamefont {Perdew},
  \citenamefont {Burke},\ and\ \citenamefont
  {Ernzerhof}}]{perdew1996generalized}%
  \BibitemOpen
  \bibfield  {author} {\bibinfo {author} {\bibfnamefont {John~P}\ \bibnamefont
  {Perdew}}, \bibinfo {author} {\bibfnamefont {Kieron}\ \bibnamefont {Burke}},
  \ and\ \bibinfo {author} {\bibfnamefont {Matthias}\ \bibnamefont
  {Ernzerhof}},\ }\bibfield  {title} {\enquote {\bibinfo {title} {Generalized
  gradient approximation made simple},}\ }\href@noop {} {\bibfield  {journal}
  {\bibinfo  {journal} {Physical review letters}\ }\textbf {\bibinfo {volume}
  {77}},\ \bibinfo {pages} {3865} (\bibinfo {year} {1996})}\BibitemShut
  {NoStop}%
\bibitem [{\citenamefont {Rappe}\ \emph {et~al.}(1990)\citenamefont {Rappe},
  \citenamefont {Rabe}, \citenamefont {Kaxiras},\ and\ \citenamefont
  {Joannopoulos}}]{rappe1990optimized}%
  \BibitemOpen
  \bibfield  {author} {\bibinfo {author} {\bibfnamefont {Andrew~M}\
  \bibnamefont {Rappe}}, \bibinfo {author} {\bibfnamefont {Karin~M}\
  \bibnamefont {Rabe}}, \bibinfo {author} {\bibfnamefont {Efthimios}\
  \bibnamefont {Kaxiras}}, \ and\ \bibinfo {author} {\bibfnamefont
  {JD}~\bibnamefont {Joannopoulos}},\ }\bibfield  {title} {\enquote {\bibinfo
  {title} {Optimized pseudopotentials},}\ }\href@noop {} {\bibfield  {journal}
  {\bibinfo  {journal} {Physical Review B}\ }\textbf {\bibinfo {volume} {41}},\
  \bibinfo {pages} {1227} (\bibinfo {year} {1990})}\BibitemShut {NoStop}%
\bibitem [{\citenamefont {Jain}\ \emph {et~al.}(2011)\citenamefont {Jain},
  \citenamefont {Hautier}, \citenamefont {Moore}, \citenamefont {Ong},
  \citenamefont {Fischer}, \citenamefont {Mueller}, \citenamefont {Persson},\
  and\ \citenamefont {Ceder}}]{jain2011high}%
  \BibitemOpen
  \bibfield  {author} {\bibinfo {author} {\bibfnamefont {Anubhav}\ \bibnamefont
  {Jain}}, \bibinfo {author} {\bibfnamefont {Geoffroy}\ \bibnamefont
  {Hautier}}, \bibinfo {author} {\bibfnamefont {Charles~J}\ \bibnamefont
  {Moore}}, \bibinfo {author} {\bibfnamefont {Shyue~Ping}\ \bibnamefont {Ong}},
  \bibinfo {author} {\bibfnamefont {Christopher~C}\ \bibnamefont {Fischer}},
  \bibinfo {author} {\bibfnamefont {Tim}\ \bibnamefont {Mueller}}, \bibinfo
  {author} {\bibfnamefont {Kristin~A}\ \bibnamefont {Persson}}, \ and\ \bibinfo
  {author} {\bibfnamefont {Gerbrand}\ \bibnamefont {Ceder}},\ }\bibfield
  {title} {\enquote {\bibinfo {title} {A high-throughput infrastructure for
  density functional theory calculations},}\ }\href@noop {} {\bibfield
  {journal} {\bibinfo  {journal} {Computational Materials Science}\ }\textbf
  {\bibinfo {volume} {50}},\ \bibinfo {pages} {2295--2310} (\bibinfo {year}
  {2011})}\BibitemShut {NoStop}%
\bibitem [{\citenamefont {Ong}\ \emph {et~al.}(2013)\citenamefont {Ong},
  \citenamefont {Richards}, \citenamefont {Jain}, \citenamefont {Hautier},
  \citenamefont {Kocher}, \citenamefont {Cholia}, \citenamefont {Gunter},
  \citenamefont {Chevrier}, \citenamefont {Persson},\ and\ \citenamefont
  {Ceder}}]{ong2013python}%
  \BibitemOpen
  \bibfield  {author} {\bibinfo {author} {\bibfnamefont {Shyue~Ping}\
  \bibnamefont {Ong}}, \bibinfo {author} {\bibfnamefont {William~Davidson}\
  \bibnamefont {Richards}}, \bibinfo {author} {\bibfnamefont {Anubhav}\
  \bibnamefont {Jain}}, \bibinfo {author} {\bibfnamefont {Geoffroy}\
  \bibnamefont {Hautier}}, \bibinfo {author} {\bibfnamefont {Michael}\
  \bibnamefont {Kocher}}, \bibinfo {author} {\bibfnamefont {Shreyas}\
  \bibnamefont {Cholia}}, \bibinfo {author} {\bibfnamefont {Dan}\ \bibnamefont
  {Gunter}}, \bibinfo {author} {\bibfnamefont {Vincent~L}\ \bibnamefont
  {Chevrier}}, \bibinfo {author} {\bibfnamefont {Kristin~A}\ \bibnamefont
  {Persson}}, \ and\ \bibinfo {author} {\bibfnamefont {Gerbrand}\ \bibnamefont
  {Ceder}},\ }\bibfield  {title} {\enquote {\bibinfo {title} {Python materials
  genomics (pymatgen): A robust, open-source python library for materials
  analysis},}\ }\href@noop {} {\bibfield  {journal} {\bibinfo  {journal}
  {Computational Materials Science}\ }\textbf {\bibinfo {volume} {68}},\
  \bibinfo {pages} {314--319} (\bibinfo {year} {2013})}\BibitemShut {NoStop}%
\bibitem [{\citenamefont {Ranmohotti}\ \emph {et~al.}(2015)\citenamefont
  {Ranmohotti}, \citenamefont {Djieutedjeu}, \citenamefont {Lopez},
  \citenamefont {Page}, \citenamefont {Haldolaarachchige}, \citenamefont {Chi},
  \citenamefont {Sahoo}, \citenamefont {Uher}, \citenamefont {Young},\ and\
  \citenamefont {Poudeu}}]{ranmohotti_coexistence_2015}%
  \BibitemOpen
  \bibfield  {author} {\bibinfo {author} {\bibfnamefont {Kulugammana G.~S.}\
  \bibnamefont {Ranmohotti}}, \bibinfo {author} {\bibfnamefont {Honore}\
  \bibnamefont {Djieutedjeu}}, \bibinfo {author} {\bibfnamefont {Juan}\
  \bibnamefont {Lopez}}, \bibinfo {author} {\bibfnamefont {Alexander}\
  \bibnamefont {Page}}, \bibinfo {author} {\bibfnamefont {Neel}\ \bibnamefont
  {Haldolaarachchige}}, \bibinfo {author} {\bibfnamefont {Hang}\ \bibnamefont
  {Chi}}, \bibinfo {author} {\bibfnamefont {Pranati}\ \bibnamefont {Sahoo}},
  \bibinfo {author} {\bibfnamefont {Ctirad}\ \bibnamefont {Uher}}, \bibinfo
  {author} {\bibfnamefont {David}\ \bibnamefont {Young}}, \ and\ \bibinfo
  {author} {\bibfnamefont {Pierre F.~P.}\ \bibnamefont {Poudeu}},\ }\bibfield
  {title} {\enquote {\bibinfo {title} {Coexistence of {High}-{Tc}
  {Ferromagnetism} and n-{Type} {Electrical} {Conductivity} in {FeBi}2se4},}\
  }\href {\doibase 10.1021/ja5084255} {\bibfield  {journal} {\bibinfo
  {journal} {Journal of the American Chemical Society}\ }\textbf {\bibinfo
  {volume} {137}},\ \bibinfo {pages} {691--698} (\bibinfo {year}
  {2015})}\BibitemShut {NoStop}%
\bibitem [{sup()}]{supplement}%
  \BibitemOpen
  \href@noop {} {}\bibinfo {note} {See supplemental material for
  crystallographic information files of potentially stable
  compounds.}\BibitemShut {Stop}%
\bibitem [{\citenamefont {Stevanovi{\'c}}\ \emph {et~al.}(2012)\citenamefont
  {Stevanovi{\'c}}, \citenamefont {Lany}, \citenamefont {Zhang},\ and\
  \citenamefont {Zunger}}]{stevanovic2012correcting}%
  \BibitemOpen
  \bibfield  {author} {\bibinfo {author} {\bibfnamefont {Vladan}\ \bibnamefont
  {Stevanovi{\'c}}}, \bibinfo {author} {\bibfnamefont {Stephan}\ \bibnamefont
  {Lany}}, \bibinfo {author} {\bibfnamefont {Xiuwen}\ \bibnamefont {Zhang}}, \
  and\ \bibinfo {author} {\bibfnamefont {Alex}\ \bibnamefont {Zunger}},\
  }\bibfield  {title} {\enquote {\bibinfo {title} {Correcting density
  functional theory for accurate predictions of compound enthalpies of
  formation: Fitted elemental-phase reference energies},}\ }\href@noop {}
  {\bibfield  {journal} {\bibinfo  {journal} {Physical Review B}\ }\textbf
  {\bibinfo {volume} {85}},\ \bibinfo {pages} {115104} (\bibinfo {year}
  {2012})}\BibitemShut {NoStop}%
\bibitem [{\citenamefont {Persson}\ \emph {et~al.}(2005)\citenamefont
  {Persson}, \citenamefont {Zhao}, \citenamefont {Lany},\ and\ \citenamefont
  {Zunger}}]{persson2005n}%
  \BibitemOpen
  \bibfield  {author} {\bibinfo {author} {\bibfnamefont {Clas}\ \bibnamefont
  {Persson}}, \bibinfo {author} {\bibfnamefont {Yu-Jun}\ \bibnamefont {Zhao}},
  \bibinfo {author} {\bibfnamefont {Stephan}\ \bibnamefont {Lany}}, \ and\
  \bibinfo {author} {\bibfnamefont {Alex}\ \bibnamefont {Zunger}},\ }\bibfield
  {title} {\enquote {\bibinfo {title} {n-type doping of cuin se 2 and cuga se
  2},}\ }\href@noop {} {\bibfield  {journal} {\bibinfo  {journal} {Physical
  Review B}\ }\textbf {\bibinfo {volume} {72}},\ \bibinfo {pages} {035211}
  (\bibinfo {year} {2005})}\BibitemShut {NoStop}%
\bibitem [{\citenamefont {Lafond}\ \emph {et~al.}(1999)\citenamefont {Lafond},
  \citenamefont {Cody}, \citenamefont {Mo{\"e}lo}, \citenamefont {Meerschaut},\
  and\ \citenamefont {Rouxel}}]{lafond1999magnetic}%
  \BibitemOpen
  \bibfield  {author} {\bibinfo {author} {\bibfnamefont {A}~\bibnamefont
  {Lafond}}, \bibinfo {author} {\bibfnamefont {JA}~\bibnamefont {Cody}},
  \bibinfo {author} {\bibfnamefont {Y}~\bibnamefont {Mo{\"e}lo}}, \bibinfo
  {author} {\bibfnamefont {A}~\bibnamefont {Meerschaut}}, \ and\ \bibinfo
  {author} {\bibfnamefont {J}~\bibnamefont {Rouxel}},\ }\bibfield  {title}
  {\enquote {\bibinfo {title} {Magnetic behavior in the incommensurate phase~
  lacrs 3:: effect of the partial substitution of cr by ti on the magnetic
  properties},}\ }\href@noop {} {\bibfield  {journal} {\bibinfo  {journal}
  {Journal of magnetism and magnetic materials}\ }\textbf {\bibinfo {volume}
  {192}},\ \bibinfo {pages} {490--498} (\bibinfo {year} {1999})}\BibitemShut
  {NoStop}%
\bibitem [{\citenamefont {Rau}\ \emph {et~al.}(1973)\citenamefont {Rau},
  \citenamefont {Kutty},\ and\ \citenamefont {De~Carvalho}}]{rau1973high}%
  \BibitemOpen
  \bibfield  {author} {\bibinfo {author} {\bibfnamefont {H\_}\ \bibnamefont
  {Rau}}, \bibinfo {author} {\bibfnamefont {TRN}\ \bibnamefont {Kutty}}, \ and\
  \bibinfo {author} {\bibfnamefont {JRF~Guedes}\ \bibnamefont {De~Carvalho}},\
  }\bibfield  {title} {\enquote {\bibinfo {title} {High temperature saturated
  vapour pressure of sulphur and the estimation of its critical quantities},}\
  }\href@noop {} {\bibfield  {journal} {\bibinfo  {journal} {The Journal of
  Chemical Thermodynamics}\ }\textbf {\bibinfo {volume} {5}},\ \bibinfo {pages}
  {291--302} (\bibinfo {year} {1973})}\BibitemShut {NoStop}%
\bibitem [{\citenamefont {Klimova}\ \emph {et~al.}(2006)\citenamefont
  {Klimova}, \citenamefont {Ananichev}, \citenamefont {Demidov},\ and\
  \citenamefont {Blinov}}]{klimova2006investigation}%
  \BibitemOpen
  \bibfield  {author} {\bibinfo {author} {\bibfnamefont {AM}~\bibnamefont
  {Klimova}}, \bibinfo {author} {\bibfnamefont {VA}~\bibnamefont {Ananichev}},
  \bibinfo {author} {\bibfnamefont {AI}~\bibnamefont {Demidov}}, \ and\
  \bibinfo {author} {\bibfnamefont {LN}~\bibnamefont {Blinov}},\ }\bibfield
  {title} {\enquote {\bibinfo {title} {Investigation of the saturated vapor
  pressure of selenium and indium sesquiselenide in0. 4se0. 6},}\ }\href@noop
  {} {\bibfield  {journal} {\bibinfo  {journal} {Glass Physics and Chemistry}\
  }\textbf {\bibinfo {volume} {32}},\ \bibinfo {pages} {436--438} (\bibinfo
  {year} {2006})}\BibitemShut {NoStop}%
\bibitem [{\citenamefont {Baker}(1968)}]{baker1968vapour}%
  \BibitemOpen
  \bibfield  {author} {\bibinfo {author} {\bibfnamefont {EH}~\bibnamefont
  {Baker}},\ }\bibfield  {title} {\enquote {\bibinfo {title} {The vapour
  pressure and resistivity of selenium at high temperatures},}\ }\href@noop {}
  {\bibfield  {journal} {\bibinfo  {journal} {Journal of the Chemical Society
  A: Inorganic, Physical, Theoretical}\ ,\ \bibinfo {pages} {1089--1092}}
  (\bibinfo {year} {1968})}\BibitemShut {NoStop}%
\bibitem [{\citenamefont {Tsai}\ \emph {et~al.}(1995)\citenamefont {Tsai},
  \citenamefont {Wang},\ and\ \citenamefont {Hon}}]{JACE:JACE1185}%
  \BibitemOpen
  \bibfield  {author} {\bibinfo {author} {\bibfnamefont {Ming-Shyong}\
  \bibnamefont {Tsai}}, \bibinfo {author} {\bibfnamefont {Li-Hsing}\
  \bibnamefont {Wang}}, \ and\ \bibinfo {author} {\bibfnamefont {Min-Hsiung}\
  \bibnamefont {Hon}},\ }\bibfield  {title} {\enquote {\bibinfo {title}
  {Formation of calcium lanthanum sulfide by carbonate coprecipitation and cs2
  sulfurization},}\ }\href {\doibase 10.1111/j.1151-2916.1995.tb08467.x}
  {\bibfield  {journal} {\bibinfo  {journal} {Journal of the American Ceramic
  Society}\ }\textbf {\bibinfo {volume} {78}},\ \bibinfo {pages} {1185--1190}
  (\bibinfo {year} {1995})}\BibitemShut {NoStop}%
\bibitem [{\citenamefont {Saad}\ \emph {et~al.}(2010)\citenamefont {Saad},
  \citenamefont {Kuzuya}, \citenamefont {Hirai},\ and\ \citenamefont
  {Ohta}}]{saad2010synthesis}%
  \BibitemOpen
  \bibfield  {author} {\bibinfo {author} {\bibfnamefont {Omar~Massoud}\
  \bibnamefont {Saad}}, \bibinfo {author} {\bibfnamefont {Toshihiro}\
  \bibnamefont {Kuzuya}}, \bibinfo {author} {\bibfnamefont {Shinji}\
  \bibnamefont {Hirai}}, \ and\ \bibinfo {author} {\bibfnamefont {Michihiro}\
  \bibnamefont {Ohta}},\ }\bibfield  {title} {\enquote {\bibinfo {title}
  {Synthesis of lncus2 (ln= ce, pr, nd, sm, gd, and tb) powder by polymerized
  complex method and cs2 gas sulfurization},}\ }\href@noop {} {\bibfield
  {journal} {\bibinfo  {journal} {Materials transactions}\ }\textbf {\bibinfo
  {volume} {51}},\ \bibinfo {pages} {2289--2293} (\bibinfo {year}
  {2010})}\BibitemShut {NoStop}%
\bibitem [{\citenamefont {Hirai}\ \emph {et~al.}(1998)\citenamefont {Hirai},
  \citenamefont {Shimakage}, \citenamefont {Saitou}, \citenamefont {Nishimura},
  \citenamefont {Uemura}, \citenamefont {Mitomo},\ and\ \citenamefont
  {Brewer}}]{hirai1998synthesis}%
  \BibitemOpen
  \bibfield  {author} {\bibinfo {author} {\bibfnamefont {Shinji}\ \bibnamefont
  {Hirai}}, \bibinfo {author} {\bibfnamefont {Kazuyoshi}\ \bibnamefont
  {Shimakage}}, \bibinfo {author} {\bibfnamefont {Yasushi}\ \bibnamefont
  {Saitou}}, \bibinfo {author} {\bibfnamefont {Toshiyuki}\ \bibnamefont
  {Nishimura}}, \bibinfo {author} {\bibfnamefont {Yoichiro}\ \bibnamefont
  {Uemura}}, \bibinfo {author} {\bibfnamefont {Mamoru}\ \bibnamefont {Mitomo}},
  \ and\ \bibinfo {author} {\bibfnamefont {Leo}\ \bibnamefont {Brewer}},\
  }\bibfield  {title} {\enquote {\bibinfo {title} {Synthesis and sintering of
  cerium (iii) sulfide powders},}\ }\href@noop {} {\bibfield  {journal}
  {\bibinfo  {journal} {Journal of the American Ceramic Society}\ }\textbf
  {\bibinfo {volume} {81}},\ \bibinfo {pages} {145--151} (\bibinfo {year}
  {1998})}\BibitemShut {NoStop}%
\bibitem [{\citenamefont {Coelho}(2004)}]{coelho2004topas}%
  \BibitemOpen
  \bibfield  {author} {\bibinfo {author} {\bibfnamefont {Alan~A}\ \bibnamefont
  {Coelho}},\ }\bibfield  {title} {\enquote {\bibinfo {title}
  {Topas-academic},}\ }\href@noop {} {\bibfield  {journal} {\bibinfo  {journal}
  {A Computer Programme for Rietveld Analysis}\ } (\bibinfo {year}
  {2004})}\BibitemShut {NoStop}%
\bibitem [{\citenamefont {Shoemaker}\ \emph {et~al.}(2012)\citenamefont
  {Shoemaker}, \citenamefont {Chung}, \citenamefont {Mitchell}, \citenamefont
  {Bray}, \citenamefont {Soderholm}, \citenamefont {Chupas},\ and\
  \citenamefont {Kanatzidis}}]{shoemaker2012understanding}%
  \BibitemOpen
  \bibfield  {author} {\bibinfo {author} {\bibfnamefont {Daniel~P.}\
  \bibnamefont {Shoemaker}}, \bibinfo {author} {\bibfnamefont {Duck~Young}\
  \bibnamefont {Chung}}, \bibinfo {author} {\bibfnamefont {J.~F.}\ \bibnamefont
  {Mitchell}}, \bibinfo {author} {\bibfnamefont {Travis~H.}\ \bibnamefont
  {Bray}}, \bibinfo {author} {\bibfnamefont {L.}~\bibnamefont {Soderholm}},
  \bibinfo {author} {\bibfnamefont {Peter~J.}\ \bibnamefont {Chupas}}, \ and\
  \bibinfo {author} {\bibfnamefont {Mercouri~G.}\ \bibnamefont {Kanatzidis}},\
  }\bibfield  {title} {\enquote {\bibinfo {title} {Understanding fluxes as
  media for directed synthesis: In situ local structure of molten potassium
  polysulfides},}\ }\href {\doibase 10.1021/ja303047e} {\bibfield  {journal}
  {\bibinfo  {journal} {J. Am. Chem. Soc.}\ }\textbf {\bibinfo {volume}
  {134}},\ \bibinfo {pages} {9456--9463} (\bibinfo {year} {2012})}\BibitemShut
  {NoStop}%
\bibitem [{\citenamefont {Yee}\ \emph {et~al.}(2015)\citenamefont {Yee},
  \citenamefont {Birol},\ and\ \citenamefont
  {Kotliar}}]{0295-5075-111-1-17002}%
  \BibitemOpen
  \bibfield  {author} {\bibinfo {author} {\bibfnamefont {Chuck-Hou}\
  \bibnamefont {Yee}}, \bibinfo {author} {\bibfnamefont {Turan}\ \bibnamefont
  {Birol}}, \ and\ \bibinfo {author} {\bibfnamefont {Gabriel}\ \bibnamefont
  {Kotliar}},\ }\bibfield  {title} {\enquote {\bibinfo {title} {Guided design
  of copper oxysulfide superconductors},}\ }\href
  {http://stacks.iop.org/0295-5075/111/i=1/a=17002} {\bibfield  {journal}
  {\bibinfo  {journal} {EPL (Europhysics Letters)}\ }\textbf {\bibinfo {volume}
  {111}},\ \bibinfo {pages} {17002} (\bibinfo {year} {2015})}\BibitemShut
  {NoStop}%
\bibitem [{\citenamefont {Wagner}(2007)}]{wagner2007transition}%
  \BibitemOpen
  \bibfield  {author} {\bibinfo {author} {\bibfnamefont {Lucas~K}\ \bibnamefont
  {Wagner}},\ }\bibfield  {title} {\enquote {\bibinfo {title} {Transition metal
  oxides using quantum monte carlo},}\ }\href@noop {} {\bibfield  {journal}
  {\bibinfo  {journal} {Journal of Physics: Condensed Matter}\ }\textbf
  {\bibinfo {volume} {19}},\ \bibinfo {pages} {343201} (\bibinfo {year}
  {2007})}\BibitemShut {NoStop}%
\end{thebibliography}

%

\end{document}